\documentclass{article}
\newcommand{\order}{\mathcal{O}}

\newcommand{\ave}[1]{\left \langle #1 \right \rangle}
\newcommand{\TS}[3]{\mathrm{TS}_{#2}^{#3}[#1(x)]} 
\newcommand{\diff}[2]{\frac{\mathrm{d} #1}{\mathrm{d} #2}}
\newcommand\T{\rule{0pt}{2.6ex}}
\newcommand\B{\rule[-1.2ex]{0pt}{0pt}}
\usepackage{fullpage}
\usepackage{amsmath}
\usepackage{graphicx}

\usepackage{color}

\usepackage{tikz}
\usetikzlibrary{shapes,arrows,fit,backgrounds}

\definecolor{colora}{RGB}{255,32,0}
\definecolor{colorb}{RGB}{0,115,179}
\definecolor{colorc}{RGB}{240,228,66}
\definecolor{colord}{RGB}{230,154,0}
\definecolor{colore}{RGB}{205,154,179}
\definecolor{colorf}{RGB}{90,179,230}

\definecolor{S}{RGB}{0,154,128}
\definecolor{I}{RGB}{255,32,0}
\definecolor{R}{RGB}{230,154,0} 

\tikzset{box/.style={draw, minimum width = 1.cm, minimum height = 1.cm, text width=1cm, text centered, fill=white},
slantedbox/.style={draw,trapezium,trapezium left angle=70,trapezium right angle=-70, fill=colora!5, label={[shift={(-5ex,-3.5ex)}]top right corner:#1}}
}

\begin{document}
\title{Complex Contagions and hybrid phase transitions in unclustered
  and clustered random networks} 
\author{Joel C. Miller}

\maketitle

\begin{abstract}
A complex contagion is an infectious process in which individuals may require multiple transmissions before changing state.  These are used to model behaviors if an individual only adopts a particular behavior after perceiving a consensus among others.  We may think of individuals as beginning inactive and becoming active once they are contacted by a sufficient number of active partners.  These have been studied in a number of cases, but analytic models for the dynamic spread of complex contagions are typically complex.  Here we study the dynamics of the Watts Threshold Model (WTM) assuming that transmission occurs in continuous time as a Poisson process, or in discrete time where individuals transmit to all partners in the time step following their infection.  We adapt techniques developed for infectious disease modeling to develop an analyze analytic models for the dynamics of the WTM in Configuration Model networks and a class of random clustered (triangle-based) networks.  The resulting model is relatively simple and compact.  We use it to gain insights into the dynamics of the contagion.  Taking the infinite population limit, we derive conditions under which cascades happen with an arbitrarily small initial proportion active, confirming a hypothesis of Watts for this case.  We also observe hybrid phase transitions when cascades are not possible for small initial conditions, but occur for large enough initial conditions.  We derive sufficient conditions for this hybrid phase transition to occur.  We show that in many cases, if the hybrid phase transition occurs, then all individuals eventually become active.  Finally, we discuss the role clustering plays in facilitating or impeding the spread and find that the hypothesis of Watts that was confirmed in Configuration Model networks does not hold in general.  This approach allows us to unify many existing disparate observations and derive new results.

\medskip

\noindent \textbf{key words:} Watts Threshold Model, Cascades, Hybrid Bifurcation, Complex Contagion, Dynamic model
\end{abstract}

\section{Introduction}

Many ``infectious'' processes spread on social contact networks.  The most studied of these are susceptible-infected-susceptible (SIS) and susceptible-infected-recovered (SIR) diseases.  For these processes, a single transmission will cause infection if the recipient is susceptible.  However, a range of behavioral patterns such as rumor spread or technology use are believed to spread as ``complex contagions''.  Here, a node in the network typically requires more than one transmission from ``active'' nodes to become ``activated''~\cite{watts:WTM}.

These have received less attention compared to infectious diseases.  However, their importance for understanding large-scale social behaviors is significant.  In understanding processes for which individuals adopt a behavior if they sense that there is a consensus in favor of that behavior, a simple SIS- or SIR-like disease model will be unable to capture the dynamics.  Experimental evidence~\cite{centola:experiment} provides support for the hypothesis that people adopt a behavior when they perceive such a consensus, and that this has an impact on the global dynamics.   Such behaviors are important for understanding how bubbles form in stock markets or how new technologies are adopted.  

We will study the Watts Threshold Model, model introduced by~\cite{watts:WTM}.  In this model, each node $u$ is assigned a threshold $r_u$ such that if it has at least $r_u$ active neighbors it becomes active.  We allow the threshold to be a random variable depending on the degree of node $u$.  If $r_u \leq 0$, then $u$ is active at the initial time.  In frequently studied cases, $r_u$ is taken to be same for all nodes or $r_u=\alpha k_u$ represents the same proportion of neighbors.  By careful choice of the rule for assigning $r_u$, we can arrive at a range of other percolation processes as special cases of the WTM~\cite{miller:percolation_uniting}, including bond, site, $k$-core, and bootstrap percolation.

We will focus most of our attention on a continuous time version of the WTM, using an asynchronous update rule.  A node $u$ receives transmission independently from each of its active neighbors at rate $\beta$.  We also consider a discrete time version in which active nodes transmit to their neighbors with probability $1$ in the time step after becoming active.  In either case, the transmission could be the transfer of an actual infectious agent, or it might simply be the observation that $v$ has adopted some new behavior.  This is sometimes referred to as a ``message'' which is passed~\cite{karrer:message,shrestha:message}, but in this study we will use the term ``transmit''.  Once node $u$ has received $r_u$ transmissions, it immediately switches to active and then begins to transmits to its neighbors.

To study this, we will adapt a compartmental modeling approach focusing on the status of edges rather than nodes.  This ``Edge-based compartmental modeling'' technique was originally developed for SIR disease spread and can be applied with a wide range of assumptions about the population or disease process~\cite{miller:volz,miller:ebcm_overview,miller:ebcm_hierarchy}.  It allows for a significant reduction in the number of equations used compared to other approaches~\cite{lindquist,ball:network_eqns,eames:pair,rattana:weighted}, in some cases reducing an infinite (or even doubly infinite) system of differential equations to an equivalent system governed by a single differential equation~\cite{miller:equivalence,rattana:pairwise}.  We will see that the same approach can give a significant simplification here as well.  Because the WTM contains many other percolation processes as special cases~\cite{miller:percolation_uniting}, this approach can be adapted to those as well.

The primary network class we study is the Configuration Model networks, a class of random networks which are determined by a degree distribution.  These networks have a negligible amount of short cycles: the existence of an edge from $u$ to $v$ and $u$ to $w$ provides no information about the length of alternate paths between $v$ and $w$.  Consequently, as long as $u$ remains inactive, knowledge about the status of $v$ provides no information about the status of $w$.  

It is generally believed that clustering enhances the spread of
complex
contagions~\cite{hackett:cascades_random_clustered,ikeda:cascade,centola:weakness},
and experimental evidence supports this~\cite{centola:experiment}.
The basis for this argument is that clustering will tend to introduce
correlations between the statuses of a node's neighbors.  Thus
a node who sees one active neighbor is more likely to see
others, and thus more likely to activate.  In contrast it is fairly
well established that clustering tends to reduce the spread of SIR
diseases, even if only
mildly~\cite{miller:random_clustered,gleeson:clustering_effect,miller:RSIcluster,kiss:cluster_comment,
  volz:clustered_result,miller:RSIcluster,melnik:unreasonable}.  Work by~\cite{hackett:cascades_random_clustered} suggests that clustering may increase or decrease the final size of WTM cascades.  We will use our model to study how clustering alters the dynamics of WTM cascades by studying a specialized model of clustered networks independently introduced by~\cite{miller:random_clustered,newman:cluster_alg}.

In our study we find hybrid phase transitions as the initial active proportion increases.  That is, the final size has a square root scaling on one side of the bifurcation point with a large jump in size at the transition.  These appear to be common in the WTM.  We discuss some of the examples we encounter.  Although these do not appear to have been seen in the WTM when initial proportion is the bifurcation parameter, they have been seen in several types of percolation that are a special case of the WTM~\cite{chalupa:bootstrap,baxter:heterogeneous_kcore,dorogovtsev:review} and a similar bifurcation exists for the WTM with average degree as the bifurcation parameter, so this result is not surprising of its own right.  We are able to find some sufficient conditions for the hybrid phase transition to occur.  In particular, there is a hybrid phase transition if $r\leq k-1$ for all nodes and neither the $r=1$ nodes nor the $r=k-1$ nodes form a giant component.  Furthermore in this scenario we anticipate that above this threshold all nodes in the giant component eventually activate, but we do not have a rigorous proof.

Related analytic work has primarily focused on the final size of cascades~\cite{gleeson:seed,hackett:cascades_random_clustered,gleeson:cascades,melnik:modularnetworks,watts:WTM}.  There has been limited work investigating dynamic spread.  In particular section III of~\cite{gleeson:cascades} (see also~\cite{melnik:modularnetworks}) uses equations that implicitly assume that once a node reaches its threshold of active neighbors there is a delay before it becomes active as and once active it immediately transmits to all of its neighbors.  Another model of~\cite{shrestha:message} allows a transmission rate $\beta$ and focused on the case of a fixed value of $r$ with a final transmission probability less than $1$.  This paper derived equations similar to those we will derive for the continuous time spread in configuration model networks, but did not analyze them in detail.  

In this paper, we demonstrate a mathematical approach that leads to a straightforward derivation of the governing equations.  We then analyze the resulting equations, introducing some new results on the behavior of cascades and unifying a range of disparate results about the WTM in random networks.

\section{The \emph{test node}}

Our derivation of the governing equations follows the ``edge-based compartmental modeling'' approach of~\cite{miller:ebcm_overview,miller:ebcm_structure,miller:ebcm_hierarchy}.  This approach is based on the observation that if the population-scale dynamics of an epidemic are deterministic, then the probability a random node has a given status is equal to the proportion of the population that has that status.  We perform a subtle shift of focus: rather than trying to calculate the proportion of the population that is in each status, we find it to be easier to calculate the probability a random node has each status.

We will consider a randomly chosen node and modify it such that it cannot transmit infection.  When first seen, this modification is often confusing, and we explain it in more detail below.  It is equivalent to the ``cavity method'' used in the message-passing approaches~\cite{karrer:message}.  The resulting model is based on probability generating functions, and so it might be more appropriately called a ``probability generating function'' method.  However we avoid this term because depending on the details of the situation modeled, the resulting equations may not be probability generating functions.  Indeed in our case, we will see functions that are similar to probability generating functions, with subtle differences.

The basic derivation can be summarized in four steps:
\begin{itemize}
\item We begin with the hypothesis that the population-scale dynamics are deterministic, but node-scale events are stochastic.

\item If the dynamics are deterministic, then the specific timing of when (or even whether) a random node $u$ is activated has negligible impact on the population-scale dynamics.  So removing $u$ from the population would have negligible impact on the dynamics.

\item Rather than removing $u$ we simply prevent it from transmitting to its neighbors, but allow it to otherwise be as before.   The resulting size is larger than would be seen if $u$ were removed, but smaller than if $u$ were not modified.  Since the difference between these is negligible, the resulting dynamics are indistinguishable from the original.

\item Preventing $u$ from transmitting has no impact on $u$'s status.  Since $u$ is a randomly chosen node, the probability it would have a given status at a given time must match the proportion of the population that would have that status at that time (because the dynamics are deterministic).  Thus calculating the probability the modified $u$ has a given status is equivalent to calculating the proportion of the population with that given status if $u$ were not modified.

\end{itemize}

We note that a careful analysis shows that the second and third points are actually unneeded.  While they are true, the key feature is that if the dynamics are deterministic then the probability an unaltered randomly chosen node $u$ has a given status equals the proportion of the population with that status, and the probability the unaltered node $u$ has a given status equals the probability that it has that status even after alteration.  Thus calculating the probability the altered node $u$ has a given status tells us the proportion of the population with each status in the original population.

We call the altered randomly chosen node $u$ a \emph{test node}.  In the configuration model case, the fact that $u$ cannot transmit to its partners means that their statuses are independent of one another.  We will find a consistency equation which gives us the probability that a random neighbor of $u$ is still inactive, which in turn allows us to determine the probability $u$ is active or inactive.

\section{The WTM in Configuration Model networks}

A Configuration Model network is generated by assigning each node a degree $k$ independently of any other degrees assigned in the population.  Each node is then given $k$ ``stubs'' and stubs are randomly paired until each stub is in exactly one pair which forms an edge.  We use $P(k,r)$ to be the probability a random node has degree $k$ and threshold $r$.  A neighbor of a random node has degree $k$ and threshold $r$ with probability $kP(k,r)/\ave{K}$ where $\ave{K}$ is the average degree.  This size bias represents the fact that the probability $v$ is a neighbor of a random node is proportional to the number of neighbors $v$ has.  

A consequence of the formation process is that a random neighbor of a node has a given degree $k$ with probability $P_n(k,r) = k P(k,r)/\ave{K}$ where $\ave{K}$ represents the average degree. We note that a neighbor's random neighbor has a given $k$ and $r$ also with probability $P(k,r)$.  We will determine the probability a random neighbor $v$ of the test node $u$ has not yet transmitted  to $u$ in terms of the probability that $v$'s random neighbors have not transmitted to $u$.  Since the neighbors of $v$ (other than $u$) are chosen from the same distribution as $v$, this will be the step that leads us to a consistency equation for the probability $v$ is still inactive in terms of the probability that neighbors of $v$ are still inactive.

\subsection{Continuous time}
We begin by studying a continuous time model.  In this model an active neighbor $v$ of $u$ (with threshold $r_u$) transmits to $u$ at rate $\beta$ (as a Poisson process).  Once the $r_u$th neighbor of $u$ has transmitted to $u$, $u$ immediately activates.  Once a node (other than the test node) activates it begins transmitting immediately.  Note that we do not count repeated transmissions from $v$ to $u$.  

\begin{table}
\begin{tabular}{|c | c|}
\hline\parbox{0.2\textwidth}{\textbf{Variable}} & \parbox{0.7\textwidth}{\textbf{Description}}\\
\hline\hline
\parbox{0.2\textwidth}{$P(k,r)$}& \parbox{0.7\textwidth}{\T{}The probability that a random node has degree $k$ and threshold $r$.\B{}}\\ \hline 
\parbox{0.2\textwidth}{$\beta$}& \parbox{0.7\textwidth}{\T{}Transmission rate from active node to partner.\B{}}\\ \hline 
\parbox{0.2\textwidth}{$Q(t)$} & \parbox{0.7\textwidth}{\T{}The probability a test node $u$ is still quiescent (inactive).\B{}}\\ \hline 
\parbox{0.2\textwidth}{$A(t)$} & \parbox{0.7\textwidth}{\T{}The probability a test node $u$ is active.\B{}}\\ \hline
\parbox{0.2\textwidth}{$\phi_Q(t)$}& \parbox{0.7\textwidth}{\T{}The probability a random neighbor of a test node $u$ is still quiescent.\B{}}\\ \hline 
\parbox{0.2\textwidth}{$\phi_A(t)$}& \parbox{0.7\textwidth}{\T{}The probability a random neighbor of a test node $u$ is active \emph{but has not yet transmitted to $u$}.\B{}}\\ \hline 
\parbox{0.2\textwidth}{$\theta(t)=\phi_Q(t)+\phi_A(t)$} & \parbox{0.7\textwidth}{\T{}The probability a random neighbor of the test node $u$ has not transmitted to $u$.\B{}}\\ \hline
\parbox{0.2\textwidth}{$1-\theta(t)$}& \parbox{0.7\textwidth}{\T{}The probability a random neighbor of the test node $u$ has transmitted to $u$.\B{}}\\ \hline 
\parbox{0.2\textwidth}{$\TS{f}{x_0}{n}|_{x=x_1}$}& \parbox{0.7\textwidth}{\T{}$\sum_{j=0}^n f^{(j)}(x_0)(x-x_0)^j/j!$  the Taylor Series of $f(x)$ centered at $x_0$, evaluated at $x_1$ and truncated at the $n$th term.\B{}}\\ \hline 
\parbox{0.2\textwidth}{$\psi_{r,CM}(x)$}& \parbox{0.7\textwidth}{\T{}$ \sum_{k=0}^\infty P(k,r) x^k $ a function related to probability generating functions which will be used to determine the proportion of the population that is both quiescent and has threshold $r$.\B{}}\\ \hline 
\parbox{0.2\textwidth}{$\ave{K}= \sum_{k,r} k P(k,r)$}& \parbox{0.7\textwidth}{\T{}The average degree of nodes in the population\B{}}\\ \hline 
\end{tabular}
\caption{The variables used in the continuous time model for transmission in a configuration model network}
\label{tab:variables}
\end{table}

We let $u$ be the randomly chosen test node and $v$ be a random neighbor of $u$.  If $w$ is another neighbor of $u$, the status of $v$ is independent of the status of $w$ because $u$ is prevented from transmitting and thus information about one cannot pass to the other.  

\begin{figure}
\begin{tikzpicture}
\node [box] (phiQ) at (-1.5,0) {$\phi_Q$};
\node [box] (phiA) at (1.5,0) {$\phi_A$};
\node [box] (omt) at (1.5,-4.5) {$1-\theta$};
\begin{pgfonlayer}{background}
\node [slantedbox=$\theta$, fit = (phiQ)(phiA) , minimum height = 2cm] at (0,0) (Btheta) {};
\node [slantedbox={}, fit = (omt), minimum height = 2cm] at (1.5,-4.5) (OMT) {};
\end{pgfonlayer}
\path [->] (phiQ) edge node {} (phiA);
\path [->, right] (phiA) edge node {$\beta \phi_A$} (omt);
\end{tikzpicture}
\hfill 
\begin{tikzpicture}
\node [box] (Q) at (0,0) {$Q$};
\node [box] (A) at (0,-4.5) {$A$};
\path [->] (Q) edge node {} (A);
\end{tikzpicture}
\hfill~
\caption{(left) A flow diagram showing the transitions between the possible states for $v$: From being quiescent (with probability $\phi_Q$) to being active but having not yet transmitted to $u$ (with probability $\phi_A$) to having transmitted (with probability $1-\theta$).  (right) A flow diagram showing the transitions for possible states for $u$: from quiescent to active.}
\label{fig:phi_flow}
\end{figure}
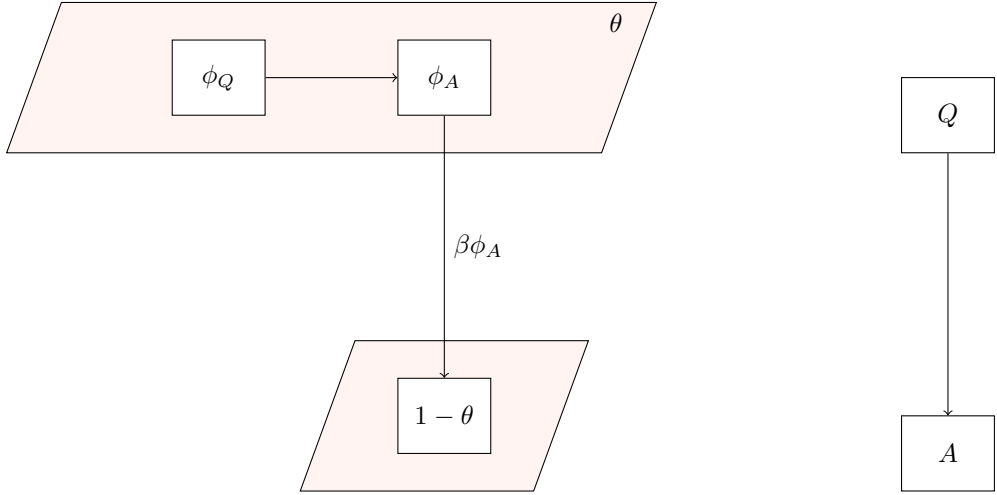

We define $\theta(t)$ to be the probability that $v$ has not yet transmitted to $u$.  We break $\theta(t)$ into two parts: $\phi_Q(t)$, the probability $v$ has not yet transmitted  and is quiescent, and $\phi_A(t)$, the probability $v$ has not yet transmitted to $u$ but is active.  Then $\theta(t)=\phi_Q(t)+\phi_A(t)$ and $1-\theta(t)$ is the probability $v$ has transmitted to $u$.  We demonstrate the flow of probability between these compartments in figure~\ref{fig:phi_flow}.

The probability $u$ is still quiescent is the probability that fewer than $r$ neighbors have transmitted to $u$.  Taking $m$ to be the number of transmissions $u$ has received, and summing over all possible values of $k$, $r$, and $m<r$ we get
\begin{align*}
Q(t) &= \sum_k \sum_{r>0} \sum_{m=0}^{r-1} P(k,r) \binom{k}{m} \theta(t)^{k-m} (1-\theta)^m\\
&= \sum_{r>0} \sum_{m=0}^{r-1} \frac{\psi_{r,CM}^{(m)}(\theta(t))}{m!} (1-\theta(t))^m
\end{align*}
where $\psi_{r,CM}(x) = \sum_k P(k,r)x^k$.  Later when we consider networks with triangles, there will be a related function $\psi_{r,\triangle}$.  Note that the inner sum gives the first terms of the Taylor Series approximation for $\psi_{r,CM}(1)$ expanded at $\theta(t)$.  We take advantage of this to reduce the summation to
\[
Q(t) = \sum_{r>0} \TS{\psi_{R,CM}}{\theta(t)}{r-1}|_{x=1}
\]
where $\TS{f}{x_0}{n} = \sum_{j=0}^n (x-x_0)^jf^{(j)}(x_0)/j!$ is the Taylor Series of $f$ centered at $x_0$ and truncated at the $f^{(n)}$th term.

If we have $\theta(t)$, this gives us $Q(t)$, from which we can immediately find the probability $u$ is active: $A(t)=1-Q(t)$.  Thus we need an expression for $\theta(t)$, the probability a random neighbor $v$ has not transmitted to $u$.  The probability $v$ has degree $k$ and threshold $r$ is given by $k P(k,r)/\ave{K}$.  Given the degree $k$ and an $m<r$, the probability $v$ has received $m$ transmissions is $\binom{k-1}{m} \theta(t)^{k-1-m}(1-\theta(t))^m$.  The $k-1$ occurs because we know that $u$ will not transmit to $v$.  Thus the probability $v$ is quiescent is
\begin{align*}
\phi_Q(t) &= \sum_k \sum_{r>0} \sum_{m=0}^{r-1} \frac{kP(k,r)}{\ave{K}}
\binom{k-1}{m} \theta(t)^{k-1-m}(1-\theta(t))^m\\
&= \frac{1}{\ave{K}}\sum_{r>0} \sum_{m=0}^{r-1}
\frac{\psi_{r,\text{CM}}^{(m+1)}(\theta(t))}{m!} (1-\theta(t))^m\\
&= \frac{1}{\ave{K}}\sum_{r>0} \TS{\psi'_{r,\text{CM}}}{\theta(t)}{r-1}|_{x=1} \, ,
\end{align*}
with $\phi_A(t) = \theta(t) - \phi_Q(t)$.  To find $\theta(t)$, we
simply observe that
\[
\dot{\theta} = - \beta \phi_A \, .
\]
Substituting for $\phi_A$ in terms of $\theta$, our full system of equations is
\begin{align*}
\dot{\theta} &= - \beta \left(\theta - \frac{1}{\ave{K}} \sum_{r>0}
  \TS{\psi'_{r,\text{CM}}}{\theta(t)}{r-1}|_{x=1} \right)\\
Q &= \sum_{r>0} \TS{\psi_{r,\text{CM}}}{\theta(t)}{r-1}|_{x=1}\\
A &= 1-Q \, .
\end{align*}
These equations are very similar to those of~\cite{shrestha:message}
who studied the case of a fixed value of $r$ with a final transmission
probability less than $1$.  

The single ODE for $\theta$ governs the dynamics.  As initial
conditions, we assume that a fraction $\rho$ of the population is
initially active.  This appears in the equations as setting $P(k,0)$
to be nonzero for some $k$ and reducing the value of $P(k,r)$ for
other $r$ values.  If the fraction is chosen uniformly, then each
$\psi_{r,\text{CM}}$ function will contain a factor $1-\rho$, that is
$\psi_{r,\text{CM}}(\theta|\rho) = (1-\rho)
\psi_{r,\text{CM}}(\theta|0)$ where the notation
$\psi_{r,\text{CM}}(\theta|\rho)$ denotes the value of
$\psi_{r,\text{CM}}(\theta)$ given $\rho$.  These equations predict
similar dynamics to those of section III of~\cite{gleeson:cascades}
(see also~\cite{melnik:modularnetworks}), but those equations
implicitly assume that once a node reaches its threshold of active
neighbors it becomes active as a Poisson process and once active
immediately transmits to all of its neighbors.  In contrast, here
nodes immediately become active and then transmit independently to its
neighbors as a Poisson process.  In terms of when a node $v$
transmits to a given neighbor $u$, it does not matter whether the
delay occurs after $v$ reaches its threshold but before it becomes
active or after $v$ becomes active but before it transmits.  However,
this introduces correlations in that all neighbors of a given
node receive transmission at the same time.  In populations in
which there are short cycles, these correlations can cause
correlations in the times that a single node receives
transmissions.  Thus these two models have different dynamics in
clustered networks.

Note that as $t \to \infty$, the value of $\theta$ must approach a 
constant, so $\phi_A \to 0$.  This leads to
$\theta(\infty)=\phi_Q(\infty)$ from which we can find a formula for
the final size.
\[
A(\infty) = 1- \sum_{r>0} \TS{\psi_{r,\text{CM}}}{\theta(\infty)}{r-1}|_{x=1} \, ,
\]
where
\[
\theta(\infty) = \frac{1}{\ave{K}}\sum_{r>0}
\TS{\psi'_{r,\text{CM}}}{\theta(\infty)}{r-1}|_{x=1} \, .
\]
This is equivalent to an expression found by~\cite{gleeson:seed}.
The simplest way to solve this is through iteration taking an initial
guess for $\theta$ of $\theta=1$.  We see below that this iteration corresponds
to the discrete-time version of the model.

\begin{figure}
\begin{center}
\includegraphics[width=0.45\columnwidth]{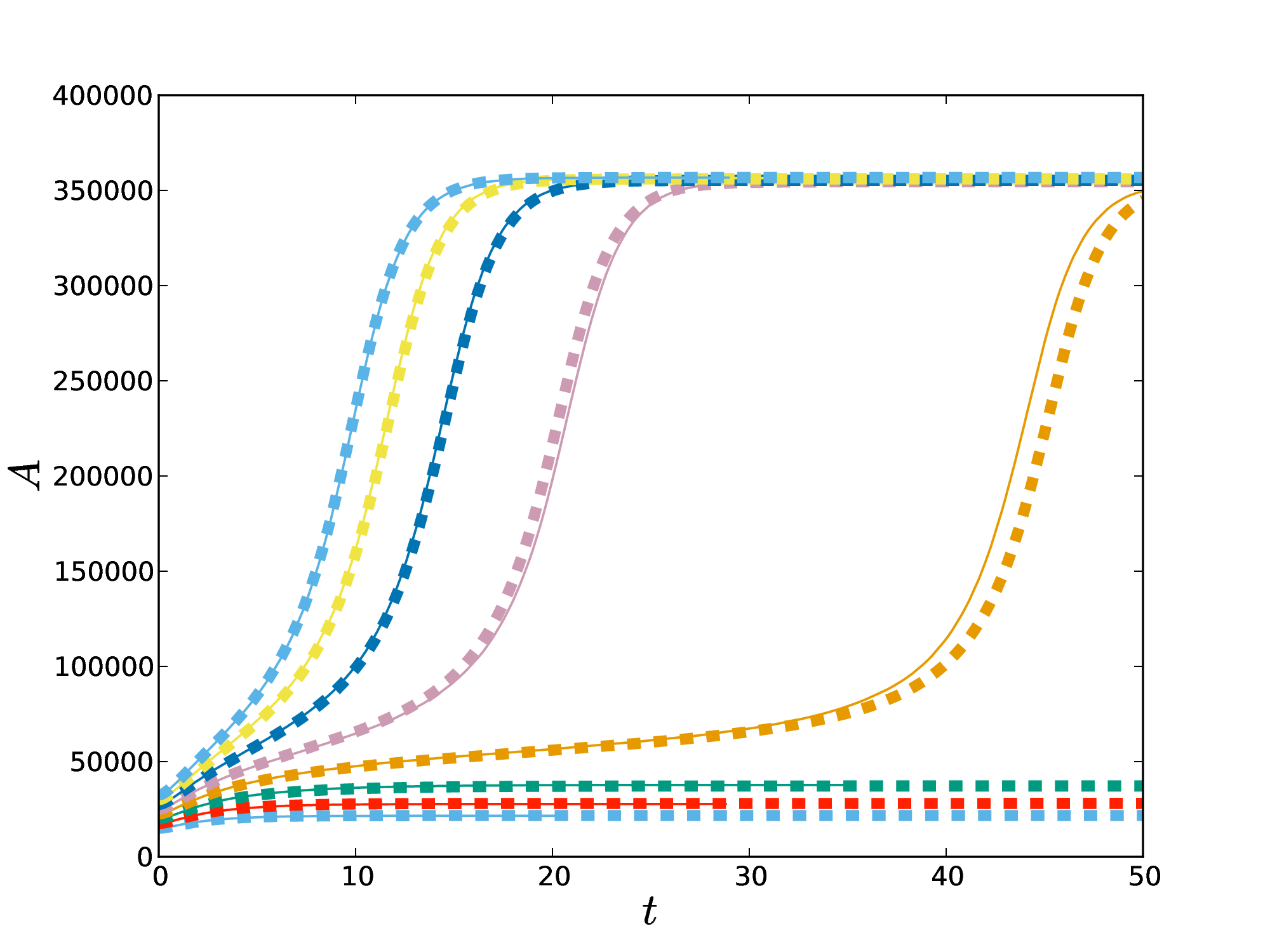}\hfill
\includegraphics[width=0.45\columnwidth]{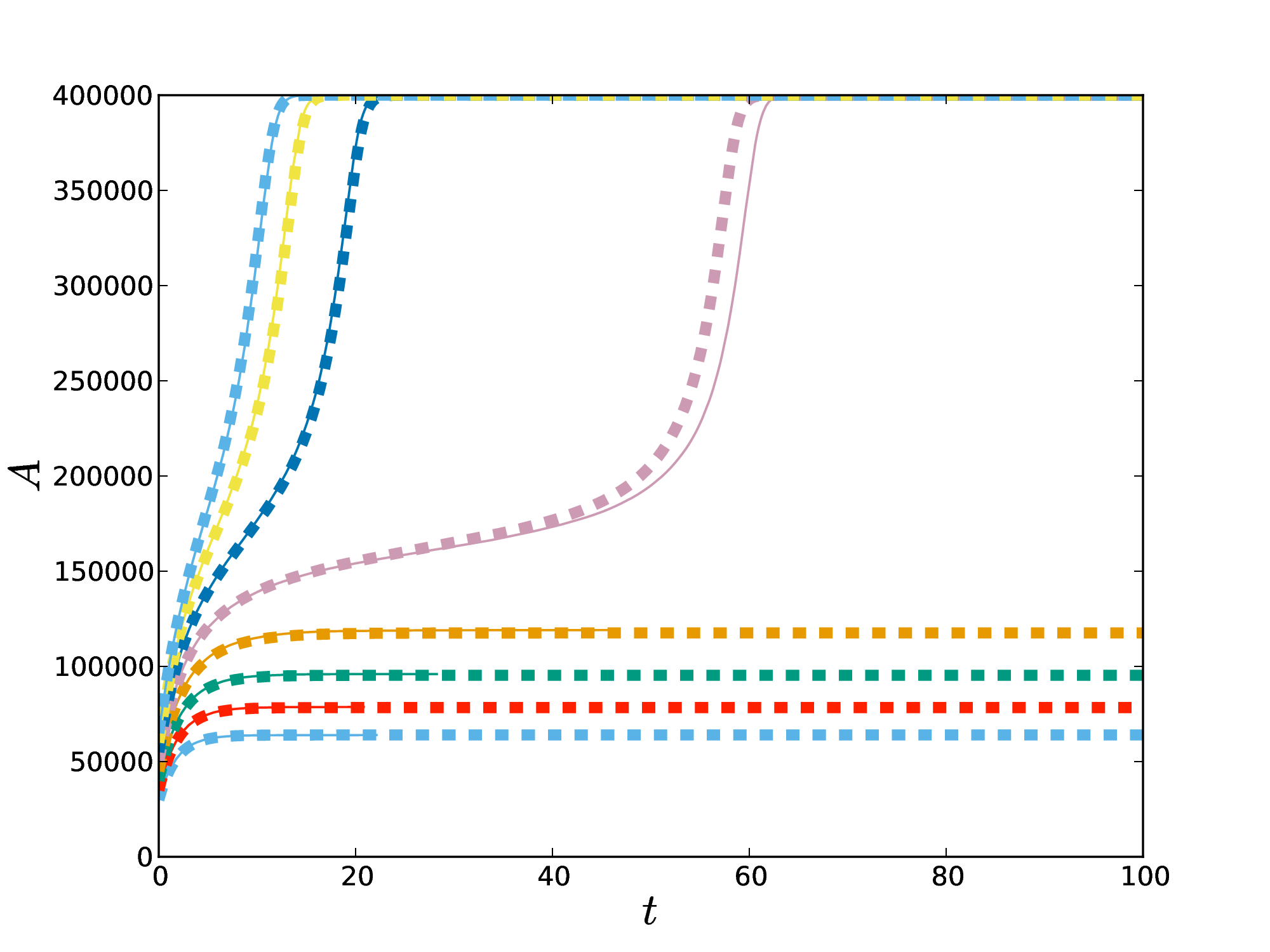}
\end{center}

\caption{The WTM in a network having nodes equally distributed between degrees $2$, $4$, and $6$.  A random fraction $\rho$ is initially active, with $\rho$ ranging from $3/80$ to $7/80$ (left) and $3/40$ to $7/40$ (right).  The other nodes either have fixed $r=2$ (left) or fixed proportion $r=k/2$ (right).  Solid curves are (Gillespie) simulations in a population of $400000$, dashed curves are theoretical predictions.  As $\rho$ increases, there is a sudden jump in the final size.  The jump is abrupt, with no initial conditions leading to an intermediate final size.  Close to this critical value, the dynamics are sensitive to stochastic effects, but elsewhere the predictions match well.  
}

\label{fig:CMfit}
\end{figure}

Figure~\ref{fig:CMfit} confirms that the model is accurate for spread in Configuration Model networks.  We are able to see a hybrid phase transition, which we discuss below.  If the initial active proportion is too small, the transmissions that occur are not enough to create significant new activations, so the numbers level off and spread halts.  At larger initial conditions, the numbers begin to level off, but a large proportion of the population are very close to reaching their activation threshold.  The large pool of almost active nodes initiates a new wave of activations and a cascade begins.  A similar behavior was observed in~\cite{melnik:multistage} where it was attributed to having multiple types of active nodes, but we see here that it is possible with just one.  It is known that at the hybrid phase transition threshold, correlation lengths diverge (in $k$-core percolation for $N \to \infty$)~\cite{goltsev:kcore}.  This long correlation magnifies stochastic effects, so close to this threshold the simulations diverge somewhat from the predictions.

\subsection{Discrete time}

We now consider a discrete time model, with synchronous updating.  Time progresses in integer units.  At each time step, if a quiescent node of threshold $r$ has at least $r$ active neighbors, it transitions to active for the following time step.  We use $t=0,1,\ldots$ to
denote the discrete time and we find that the expressions for $Q(t)$, $\phi_Q(t)$, and
$\phi_A(t)$ in terms of $\theta(t)$ remain the same as in the
continuous time model.  However, the rule for how $\theta$ updates changes
\begin{align}
\theta(t) &= \phi_Q(t-1) \nonumber \\
& = \frac{1}{\ave{K}}\sum_{r>0}
\TS{\psi_{r,\text{CM}}'}{\theta(t-1)}{r-1}|_{x=1} \, ,
\label{eqn:cobweb}
\end{align}
for $t=1,2,\ldots$ and $\theta(0)=1$.
We will be interested in
how the results change as $\rho$ changes.  To simplify our notation,
we define $f(x)$ to be the right hand side of this
\[
f(x) = \frac{1}{\ave{K}}\sum_{r>0}  \TS{\psi_{r,\text{CM}}'}{x}{r-1}|_{x=1}
\, .
\]
This has a dependence on $\rho$ through $\psi_{r,\text{CM}}$.  When we
want to make the dependence on $\rho$ explicit, we write $f(x|\rho)$.
Assuming that the initial active nodes are chosen randomly, we
have $f(x|\rho) = (1-\rho) f(x|0)$.

Note that the final size derived in the continuous dynamics case solve $\theta = f(\theta)$.  The final value $\theta(\infty)$ will be the same for this model and the continuous model.  This is because in both models any node that becomes active will eventually transmit to all of its neighbors.  The timing of those transmissions is irrelevant to the cumulative effect on the recipient.  The final state is the same.

\begin{figure}
\begin{center}
\includegraphics[width=0.45\columnwidth]{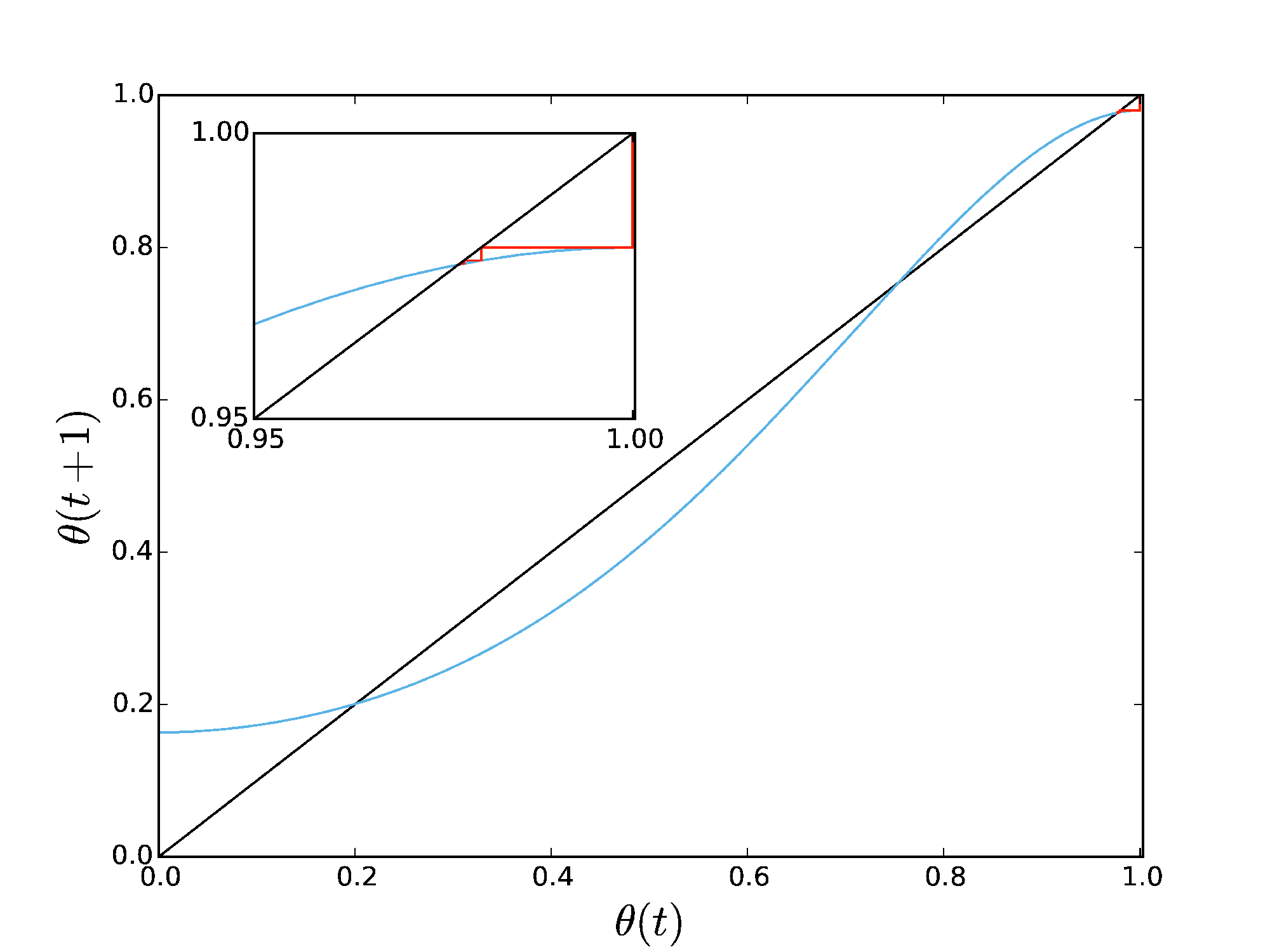}
\includegraphics[width=0.45\columnwidth]{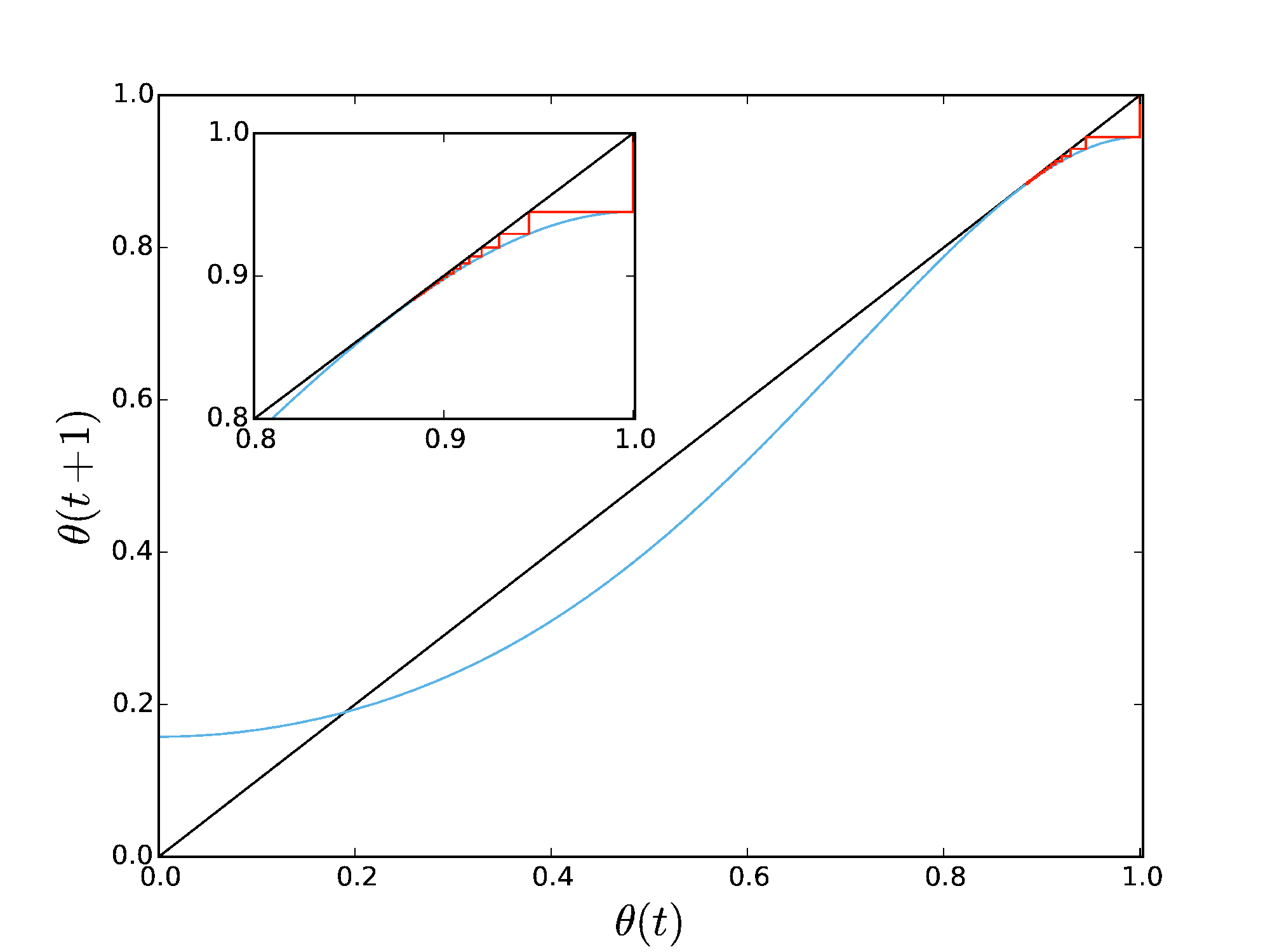}\\
\includegraphics[width=0.45\columnwidth]{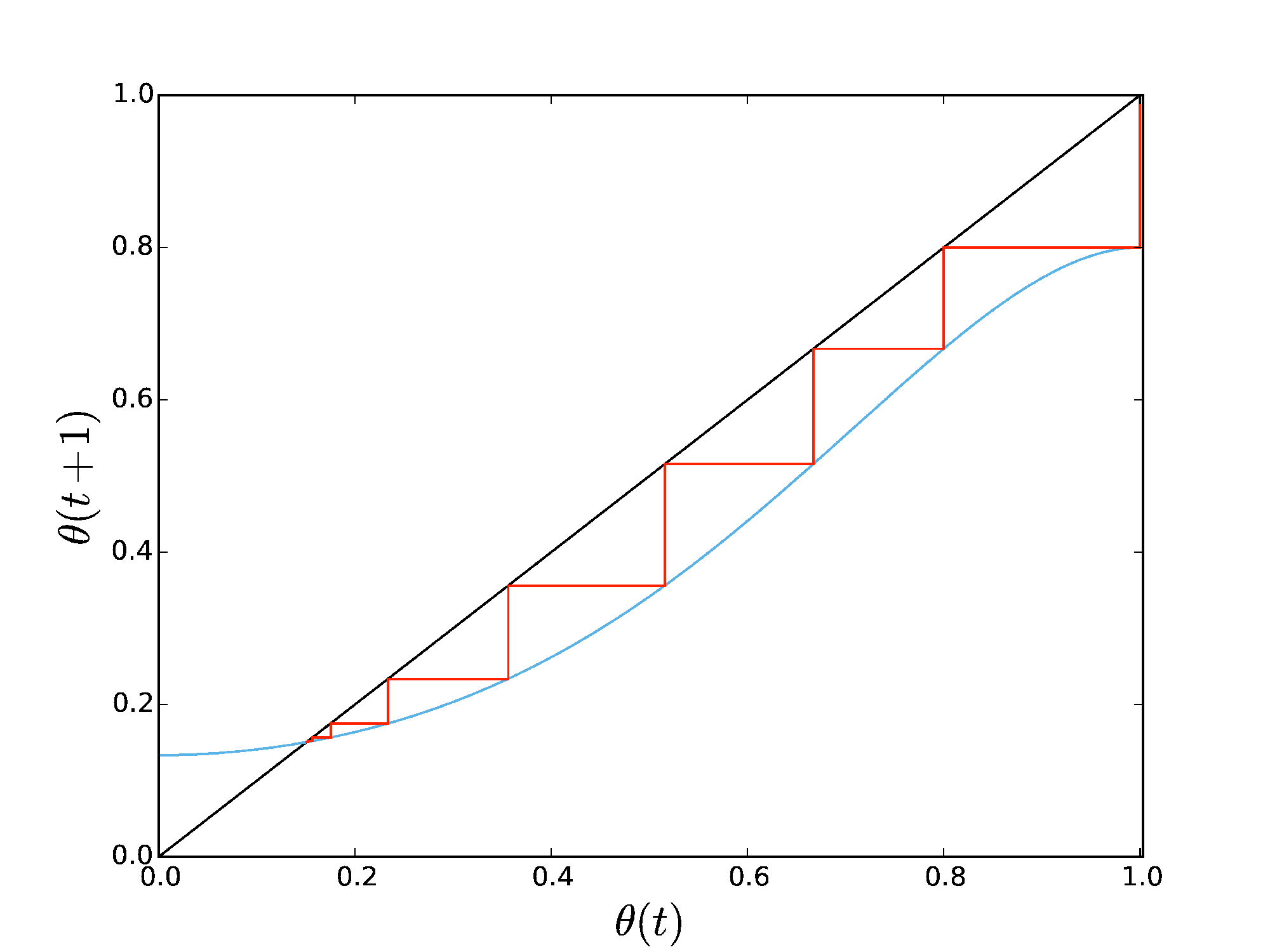}
\includegraphics[width=0.45\columnwidth]{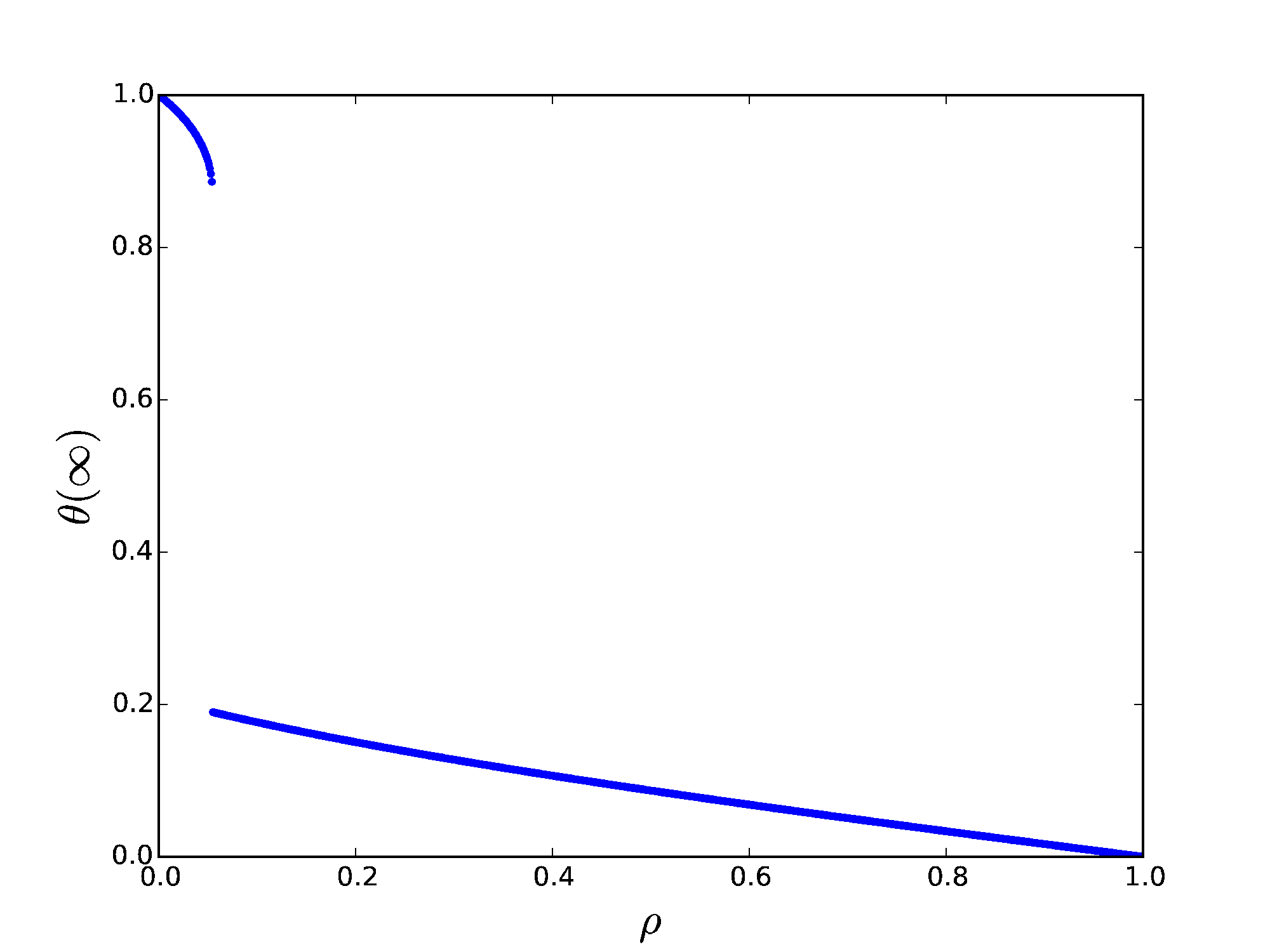}
\end{center}

\caption{Cobweb diagrams for networks with $r=2$ for all
  nodes, the degrees equally distributed between $2$, $4$, and $6$,  and a randomly chosen initially active fraction $\rho = 0.02$ (top left),
  $0.055$ (top right),
  and $0.2$ (bottom left) initially active.  Insets show detail for the
  two small  values of $\rho$.  This corresponds to the left hand plot
  of figure~\ref{fig:CMfit}.  In each case the initial $\theta$ is
  $\theta(0)=1$, and each successive value $\theta(t)$ follows from
  equation~\eqref{eqn:cobweb}.  As $\rho$ increases, there is a
  saddle-node bifurcation resulting in a discontinuous shift in
  $\theta(\infty)$.  A similar bifurcation causes the jump in final
  sizes seen in figure~\eqref{fig:CMfit}.  The final plot shows the
  values of $\theta(\infty)$ as a function of $\rho$.  Note the
  discontinuous jump with a square root scaling close to the
  threshold. }
\label{fig:cobweb}
\end{figure}

The dynamics resulting from equation~\eqref{eqn:cobweb} can be
understood through the cobweb diagrams in figure~\ref{fig:cobweb}.
Here we plot $f(\theta)$ and the diagonal line
$\theta(t+1)=\theta(t)$.  The cobweb diagram allows us to graphically
iterate $f$ to find successive values of $\theta$.

This cobweb diagram gives insight into the phase transition.  We can
use geometric arguments to show that as $\rho$ approaches the critical
value $\rho_c$ from below, the resulting value of $\theta(\infty)$ scales like
$\theta(\infty)-\theta_c \sim (\rho_c-\rho)^{1/2}$.  We can rotate and
flip the coordinates such that the diagonal $\theta(t+1)=\theta(t)$ is the
horizontal axis and then $f(\theta)$ becomes locally a parabola that crosses
the (new) horizontal axis very close to its local minimum.  By appropriately scaling
and choice of $0$, we can represent this
parabola as $x^2-c$.  The parabola crosses at $x= \pm \sqrt{c}$.
Changes in $\rho_c-\rho$ correspond to proportional changes in $c$,
with $c=0$ at the critical value of $\rho$.  The scaling of $x$ as
$\sqrt{c}$ corresponds to the scaling of $\theta(\infty)-\theta_c$.
So generically when this sort of bifurcation happens, the location of
the fixed point has the square root scaling we have identified.

We now look at this with a more analytical approach to achieve the
same result.  We use
the relation $f(\theta|\rho) = (1-\rho)f(\theta|0)$.  At
$\rho=\rho_c$, we have $\theta_c = (1-\rho_c) f(\theta_c|0)$ where
$(1-\rho_c)f'(\theta_c|0)=1$.  If we decrease $\rho$ slightly to $\rho_c
+ \hat{\rho}$ (with $\hat{\rho}<0$), we have a new limiting
$\theta(\infty)= \theta_c+\hat{\theta}$ where
\[
\theta_c+\hat{\theta} =
(1-\rho_c-\hat{\rho})f(\theta_c+\hat{\theta}|0) \, .
\]
We expand $f$ as a Taylor Series.  The obvious truncation of $f$ as
$f(\theta_c|0) + \hat{\theta}f'(\theta_c|0)$ fails because
$(1-\rho_c)f'(\theta_c|0)=1$, so the linear term
$(1-\rho_c)f'(\theta_c)\hat{\theta}$ is simply $\hat{\theta}$.  The
right hand side becomes
\begin{align*}
(1-\rho_c-\hat{\rho})f(\theta_c+\hat{\theta}|0)&=
(1-\rho_c)f(\theta_c|0) - \hat{\rho} f(\theta_c|0) \\
&\qquad + (1-\rho_c)
f'(\theta_c|0)\hat{\theta} + \cdots\\
&= \theta_c
+\hat{\theta} - \hat{\rho}f(\theta_c|0) + \cdots \, .
\end{align*}
If we do not account for the neglected terms we are left
$0=-\hat{\rho}f(\theta_c|0)$, which is wrong.  We must expand $f$ to
second order to give $0 = -\hat{\rho}f(\theta_c|0) +
(1-\rho_c)f''(\theta_c|0)\hat{\theta}^2/2$.  We conclude that
$\hat{\theta} \sim \sqrt{2\hat{\rho}
  f(\theta_c|0)/[(1-\rho_c)f''(\theta_c|0)]}$ as $\hat{\rho} \to 0^-$.
This square root dependence of $\hat{\theta}$ on $\hat{\rho}$ defines
a hybrid phase
transition~\cite{baxter:heterogeneous_kcore,dorogovtsev:review}.

\subsubsection{Alternative interpretation of discrete time model}
We can alternately interpret the discrete model in terms of
transmission chains of given lengths.  We choose a node $u$ and
a random neighbor $v$.  We prevent $u$ from transmitting to $v$ and $v$
from transmitting to its neighbors other than $u$.  We allow that the
possibility a random neighbor of $v$ transmits to $v$ might be
different from the probability $v$ transmits to $u$ (ultimately we
will see that the values are the same, but for now we allow this
possibility).  If $\theta_u$ is the probability $v$ does not transmit
to $u$ then in fact it is the probability $v$ remains quiescent
(otherwise it would eventually transmit).  If
$\theta_v$ is the probability a random other neighbor of $v$ does not
transmit to $v$, then we get
\begin{equation}
\theta_u = \frac{1}{\ave{K}}\sum_{r>0}
\TS{\psi_{r,\text{CM}}'}{\theta_v}{r-1}|_{x=1} \, .
\label{eqn:in_terms_of_neighbor}
\end{equation}
We see that if $r_v = k_v$, even if all neighbors other than $u$
transmit to $v$, that will not be enough to push $v$ over its
threshold, so $v$ will not transmit to $u$.  Note that this implies
that if every node has a threshold at least $1$ less than its degree,
then in the cobweb diagrams $f(\theta)$ goes through $0$ at
$\theta=0$.  

We can derive an equivalent expression to
equation~\eqref{eqn:in_terms_of_neighbor} for $\theta_v$ in terms of
the neighbors of $v$, and for those neighbors in terms of their neighbors.
Continuing this, we can define $\theta(t)$ to be the probability that
a random neighbor of $u$ will eventually transmit to $u$ if we discard all
nodes of distance greater than $t$ from $u$.  Then $\theta(t)$
follows the same recurrence seen before\footnote{This interpretation explains a potentially
  puzzling feature of cobweb diagrams in figure~\ref{fig:cobweb},
  namely that, for example, if we take $\theta(t-1)=0$, we find
  $\theta(t)>0$.  Previously the interpretation of this statement
  would be that if the probability a neighbor has transmitted prior to
  time $t-1$ was $1$, then the probability it has transmitted prior to
  the later time $t$ is less than $1$, which of course is nonsensical
  since the probability an event has not happened can only decrease in
  time.  This does not present a problem for the previous analysis
  because the relevant initial value is $\theta(0)=1$, so the system
  never reaches this scenario.}
\[
\theta(t) = \frac{1}{\ave{K}}\sum_{r>0}
\TS{\psi_{r,\text{CM}}'}{\theta(t-1)}{r-1}|_{x=1} \, ,
\]
reproducing equation~\eqref{eqn:cobweb}..

\subsection{Threshold condition}
In~\cite{watts:WTM} it is suggested that a global cascade can only occur when the $r=1$ nodes form a giant component:
\begin{quote}
In the context of this model, we conjecture that the required condition for a global cascade is that the subnetwork of vulnerable vertices must \emph{percolate} throughout the network as a whole, which is to say that the largest, connected vulnerable cluster must occupy a finite fraction of an infinite network.'
\end{quote}
This conjecture was made assuming the $\rho \to 0$ limit.  We will show that for Configuration Model networks this conjecture holds for the $\rho \to 0$ limit, but that hybrid phase transitions can occur if $\rho$ is large enough.  Later we will see that in the presence of clustering this conjecture is not true.

To see the cascade condition for $\rho \to 0$, consider the cobweb diagram of
figure~\ref{fig:cobweb}.  In the limit $\rho \to 0$, we find $f=1$ at
$\theta=1$.  A global cascade is possible in this limit only if
$f'(1)>1$.  Note that
$\TS{\psi_{r,\text{CM}}'}{x_0}{r-1}|_{x=1} = 
\psi_{r,\text{CM}}'(1) + \order((1-x)^r)$ because it is the first
$r-1$ terms of the Taylor Series for $\psi_{r,\text{CM}}'(1)$ centered
at $x$.  It follows that only the $r=1$ term in
equation~\eqref{eqn:cobweb} gives a nonzero contribution to the slope
of $f(\theta)$ at $\theta=1$ for $\rho \to 0$.  The leading order term
of $f'(1)$ in the $\rho \to 0$ limit is
$\psi_{1,\text{CM}}''(1)/\ave{K}$.  So only when
$\psi_{1,\text{CM}}''(1)/\ave{K}>1$ can a global cascade begin from an
arbitrarily small randomly chosen initial fraction $\rho$.  Because
$\psi_{1,\text{CM}}''(1) = \sum_k P(k,1)k(k-1)$, we see that this is a
statement about the degree and frequency of $r=1$ nodes.  In
fact it says that if a global cascade can begin from an arbitrarily
small randomly chosen initial condition then the $r=1$ nodes
form a giant connected component of the population.

We can also derive this threshold through a physical interpretation.  If the initial $\rho$ is small, then the condition for a cascade is the same as the condition that the ``infection'' travels infinitely far from an initial source.  Each active node will transmit to all of its neighbors, but until the cascade grows, we may assume that the network is tree-like.  So the transmission leads to activation iff the neighbor has $r=1$.  Thus the disease can be thought of as first exploring the largest component made up of $r=1$ nodes around the initial node.  Once it has reached all of these nodes, it can only infect $r=2$ or higher nodes if they have multiple neighbors in the initial $r=1$ component.  However, in a Configuration Model network, the probability the absence of short cycles implies that there are no such nodes unless the $r=1$ component percolates the network.  This generalizes theorem~5 of~\cite{amini:bootstrap} where the same result was proven under the assumption that the threshold $r_u$ is a function of the degree $k_u$ of $u$ (although the proof here has been less rigorous).

\subsubsection{Sufficient condition for the hybrid phase transition}
\begin{figure}
\begin{center}
\includegraphics[width=0.65\columnwidth]{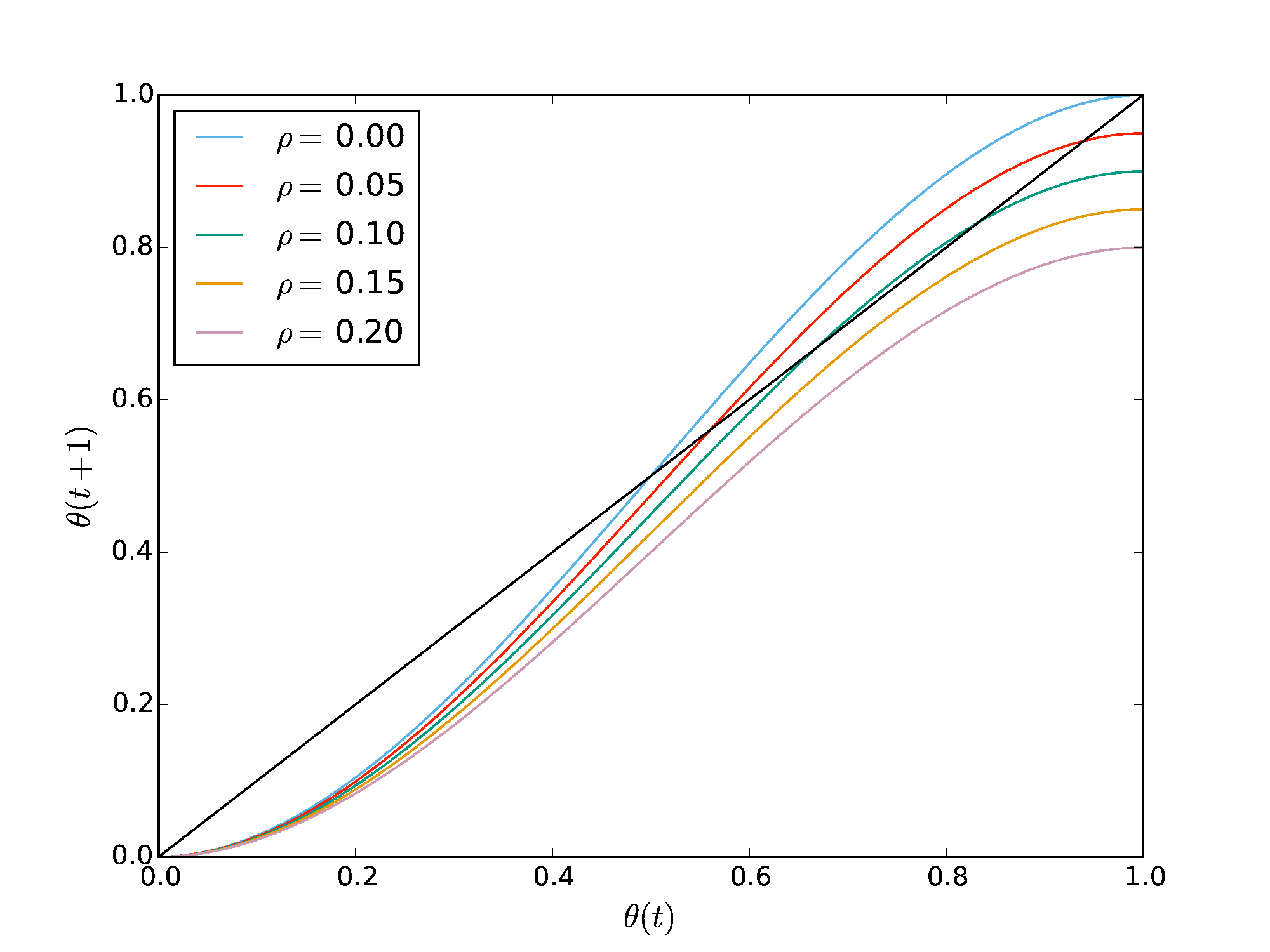}
\end{center}
\caption{Demonstrating the bifurcation when $r \leq k-1$ and neither the nodes with
  $r=k-1$ nor $r=1$ form a giant component.  Here $k=4$ and $r=2$ for
  all nodes.  When $\rho=0$, there are fixed points (places where
  $f(\theta)$ intersects the diagonal) at $0$, $1$, and an
  intermediate value.  As $\rho$ increases, this intermediate value
  moves right, while the fixed point at $1$ moves left.  Eventually
  the two meet and annihilate one another in a saddle-node
  bifurcation.  After the bifurcation, the only remaining fixed point
  is at $0$.}
\label{fig:bifurcation}
\end{figure}

We now consider the alternate question of under what condition can
there be a hybrid phase transition?  In general this appears to be a
difficult problem.  However, if we assume that every node's
threshold $r_u$ satisfies $r_u \leq k_u-1$, we can make some progress.
In this case, for $\rho=0$, there is a fixed point at $0$ and at $1$.
The slope of $f(\theta|0)$ at $\theta=0$ is given by 
\begin{align*}
\frac{1}{\ave{K}} &\sum_{r>0} \left[ \diff{}{\theta} \TS{\psi'_{r,\text{CM}}}{\theta}{r-1}|_{x=1}\right]_{\theta=0} \\
&= \frac{1}{\ave{K}} \sum_{r>0} \sum_{m=0}^{r-1}
  \left[\diff{}{\theta}
\frac{\psi_{r,\text{CM}}^{(m+1)}(\theta(t))}{m!} (1-\theta(t))^m
\right ]_{\theta=0}\\
&= \frac{1}{\ave{K}}\sum_{r>0}  \left( \psi_r^{(2)}(0)   + 
    \sum_{m=1}^{r-1}  \frac{\psi_{r,\text{CM}}^{(m+2)}(0)}{m!} -
    \frac{\psi_{r,\text{CM}}^{m+1}(0)}{(m-1)!}\right)\\
&= \frac{1}{\ave{K}} \sum_{r>0}
\frac{\psi_{r,\text{CM}}^{(r+1)}(0)}{(r-1)!}\\
&= \frac{1}{\ave{K}} \sum_{r>0} P(k=r+1,r)(r+1)r\\
&= \frac{1}{\ave{K}} \sum_k P(k,r=k-1) k(k-1) \, .
\end{align*}
If $\sum P(k,k-1)k(k-1)<\ave{K}$, then the slope at $\theta=0$ is
less than $1$.  That is, if the nodes with $r=k-1$ do not form a
giant component the slope at $\theta=0$ is less than
$1$.  At small $\theta$ the curve lies below the diagonal.\footnote{Note that in figure~\ref{fig:cobweb}, the slope of
  $f(\theta)$ at $\theta=0$ is $0$ because $P(k,k-1)=0$.  The value of
  $f$ at $\theta=0$ is nonzero because $r_u$ is not less than $k_u$
  for all nodes}  If additionally there is no cascade for arbitrarily
small $\rho$, then the derivative of the curve is less than $1$ at
$\theta=1$.  So if $r_u \leq k_u-1$ for every node $u$ and neither the
$r=1$ nodes nor the $r=k-1$ nodes form a giant component, then for
$\theta$ close to $0$, the curve is below the diagonal, while for
$\theta$ close to $1$ it is above the diagonal~\footnote{Technically
  we require that the concentration of $r=1$ or $r=k-1$ nodes must be
  strictly away from the threshold to form a giant component such that
  the inequalities are strict.  Otherwise $k=2$, \ $r=1$ for all nodes
  provides a counterexample to this result.}.  This forces the
existence of at least one additional fixed point strictly between $0$
and $1$.  Take the largest of these.  As $\rho$ increases, this fixed
point moves right, while the fixed point at $1$ moves left.
Eventually these two meet and annihilate in a saddle-node bifurcation,
resulting in a hybrid phase transition.  We see this in
figure~\ref{fig:bifurcation} for which $k=4$ and $r=2$ for all nodes.

In fact, if both of these conditions hold, then it is likely that after the
hybrid phase transition, all nodes become active.  The fixed
point at $0$ remains at $0$ for all $\rho$.  Unless there are
additional fixed points below $\theta_c$ when the bifurcation occurs,
the system will immediately move to the only remaining fixed point,
$\theta=0$.  In this case, all edges eventually transmit, and
all nodes eventually become active.

More generally, as we see in figure~\ref{fig:cobweb} we can have a
cascade even when $r_u \geq k_u$ for some nodes.  If there is no
cascade from arbitrarily small $\rho$ and there is any value of $\rho$
for which $f(\theta|\rho)$  goes below and then above the diagonal, then at a
larger value of $\rho$ this bifurcation will occur.  This is what we
see in figure~\ref{fig:cobweb}. In this case, the cascade fails to
reach the entire population.

We can generalize our sufficient condition somewhat.  If we remove the
$r\geq k-1$ nodes from the network, we are left with a new
configuration model network.  Some nodes may now have a threshold $r$
greater than or equal to their new $k-1$.  If we repeat this process
until no more nodes are removed, and the resulting core network has a giant
component, then we will satisfy the condition for a hybrid phase
transition.  

\subsubsection{Necessary conditions for hybrid phase transition}
We now show that the hybrid phase transition requires that $r>1$ for
at least some nodes.  Physically this makes sense because if $r=1$ for
all nodes, then any node in a connected component containing an
initially active node will eventually activate. We then find
another necessary condition on $f$ which is harder to translate back
into $r$.

When the saddle-node bifurcation occurs, The derivative of
$f(\theta|\rho)$ must decrease through $1$ at the bifurcation point.
This means that it must have a negative second derivative.  Since
$f(\theta|\rho) = (1-\rho)f(\theta|0)$, this means that $f(\theta|0)$
must have negative second derivative somewhere.  Taking the second
derivative of $f(\theta|0)$ gives
\begin{align*}
\sum_{r>0} \sum_{m=0}^{r-1} \frac{1}{m!}&
\Bigg(\psi_{r,\text{CM}}^{(m+3)}(\theta)(1-\theta)^m - m
  \psi_{r,\text{CM}}^{(m+2)}(\theta)(1-\theta)^{m-1}\\
& \quad+ m(m-1)
    \psi_{r,\text{CM}}^{(m+3)}(\theta)(1-\theta)^{m-2} \Bigg) \, .
\end{align*}
Because $\psi$ is defined as a summation with positive coefficients,
each derivative of $\psi$ is non-negative.  In order for the
expression above to be
negative, the middle term,
$m\psi_{r,\text{CM}}^{(m+2)}(\theta)(1-\theta)^{m-1}$ must be nonzero
for at least one value of $m$.  If $r \leq 1$ for all nodes, then when
$r<1$, the sum is empty and when $r=1$ the only term in the sum over
$m$ has $m=0$, so the term is zero.  So a hybrid phase transition in a
configuration model network requires
$r>1$ for some nodes.  With minor modifications, this shows that site
percolation (on configuration model networks) cannot exhibit this
phase transition.

Note that we can also show that if having a cascade requires $\rho$
above some positive threshold, then $f''(\theta|0)$ must change sign
between $0$ and $1$.  This is because just before the bifurcation happens,
there must be (at least) three solutions to $\theta = f(\theta|\rho)$.
The top of these three has $f'<1$, the middle has $f'>1$ and the
bottom has $f'<1$.  Since the average second derivative must be
positive in one interval and negative in the other, it must have a
sign change somewhere.  It is not clear if this leads to any simple
statements about necessary conditions for a hybrid phase transition.

\section{Triangle-based networks}

Many social networks exhibit clustering, and so the status of one neighbor $v$ of $u$ is likely to be correlated to the status of another neighbor $w$.  This assumptions of independence of neighbors breaks down.  For this case, \emph{ad hoc} approaches are able to make some progress~\cite{house:complex}.  However, in particular limits analytic results are possible.   

In~\cite{miller:random_clustered,newman:cluster_alg}, Newman and
Miller independently introduced a model of random clustered networks.
In these networks, nodes are assigned two degrees, an
independent degree $k_I$ and a triangle degree $k_\triangle$ with
probability $\hat{P}(k_I,k_\triangle)$.  a node is then given
$k_I$ regular stubs and $k_\triangle$ triangle stubs.  Regular stubs
are joined into pairs to define edges, while triangle stubs are
joined into triples to form triangles, and so a
node will have degree $k_I + 2k_\triangle$.  These networks have
clustering, but mimic some properties of ``tree-like'' networks.  This
tree-like property makes them amenable to analytic methods,
and~\cite{volz:clustered_result} showed that it is possible to apply
the edge-based compartmental modeling approach to SIR disease in such networks.

A significant weakness of this network model which must be highlighted
is that triangles in these networks do not share edges.  In fact,
locally these are a special case of ``cactus graphs'' or ``Husimi
trees''~\cite{harary:cactus}, that is, each edge is in at most one
cycle.  So although some clustering is introduced, the structure is
limited.  This can be generalized to more complicated motifs than
triangles~\cite{karrer:random_clustered,miller:random_clustered,newman:cluster_alg},
but it is still limited.  In particular, if we look at $3$--clique \cite{palla:kclique}
percolation the result will be fragmented.

\begin{figure}
\begin{center}
\includegraphics[width=0.65\columnwidth]{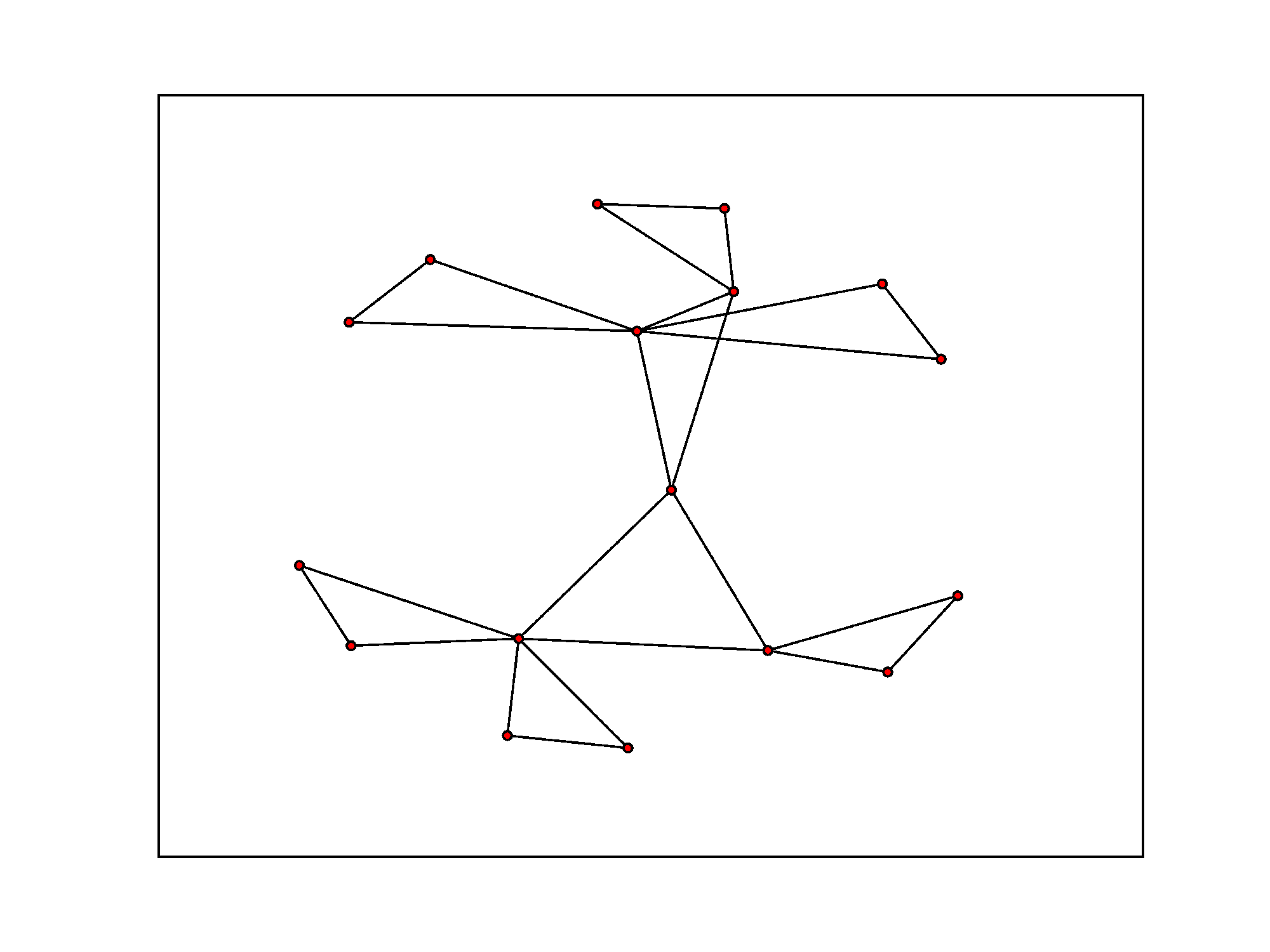}
\end{center}
\caption{A portion of a random clustered network.  For
  simplicity we choose networks with no independent edges.}
\label{fig:trianglenet}
\end{figure}

It was shown by~\cite{hackett:cascades_random_clustered} that the size of the final active nodes in the WTM can be derived analytically for this class of network.  Adapting the edge-based compartmental modeling approach leads to a model which predicts the dynamics as well.  The resulting model is relatively simple, but relies on a large number of variables.  To refine this to its purest form, we further simplify this network class, assuming that $k_I=0$ for all nodes, so there are no independent edges.  All edges appear as part of a triangle.  An example is shown in figure~\ref{fig:trianglenet}.  We use $P_\triangle(k_\triangle,r)$ to define the joint probability of being part of $k_\triangle$ triangles and having a given $r$.  We define $\psi_{r,\triangle}(x) = \sum_k P_\triangle(k_\triangle,r) x^{k_\triangle}$.

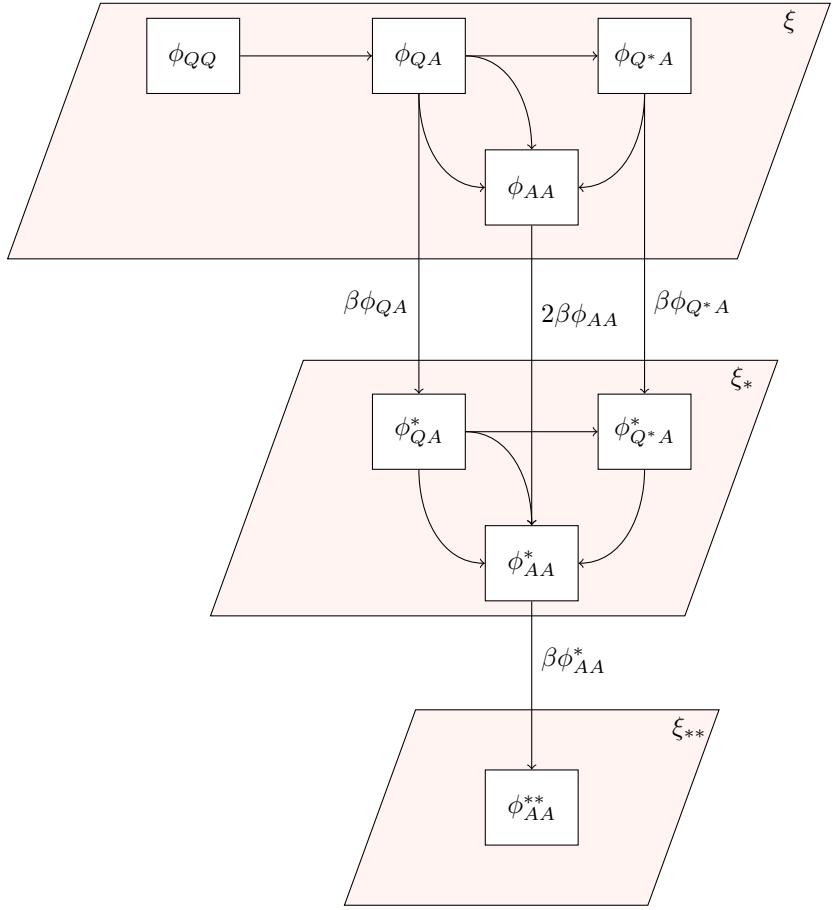
\begin{figure}
\begin{center}
\begin{tikzpicture}
\node [box] (phiQQ) at (0,0) {$\phi_{QQ}$};
\node [box] (phiQA) at (3,0) {$\phi_{QA}$};
\node [box] (phiQ*A) at (6,0) {$\phi_{Q^*A}$};
\node [box] (phiAA) at (4.5,-1.75) {$\phi_{AA}$};
\node [box] (phi*QA) at (3,-5) {$\phi^*_{QA}$};
\node [box] (phi*Q*A) at (6,-5) {$\phi^*_{Q^*A}$};
\node [box] (phi*AA) at (4.5,-6.75) {$\phi^*_{AA}$};
\node [box] (phi**AA) at (4.5,-10) {$\phi^{**}_{AA}$};
\path [->] (phiQQ) edge node {}(phiQA);
\path [->] (phiQA) edge [out=0,in=90] node {} (phiAA);
\path [->] (phiQA) edge node {} (phiQ*A);
\path [->] (phiQA) edge [out=270,in=180] node {} (phiAA);
\path [->] (phiQ*A) edge [out=270,in=0] node {} (phiAA);

\path [->, left] (phiQA) edge node[pos=0.7] {$\beta \phi_{QA}$} (phi*QA);
\path [->,right] (phiQ*A) edge node[pos=0.7]  {$\beta \phi_{Q^*A}$} (phi*Q*A);
\path [->,right] (phiAA) edge node[pos=0.3] {$2\beta\phi_{AA}$} (phi*AA);

\path [->] (phi*QA) edge [out=0,in=90] node {} (phi*AA);
\path [->] (phi*QA) edge node {} (phi*Q*A);
\path [->] (phi*QA) edge [out=270,in=180] node {} (phi*AA);
\path [->] (phi*Q*A) edge [out=270,in=0] node {} (phi*AA);

\path [->,right] (phi*AA) edge node[pos=0.35] {$\beta\phi^*_{AA}$} (phi**AA);

\begin{pgfonlayer}{background}
\node [slantedbox = $\xi$, fit = (phiQQ)(phiQA)(phiQ*A)(phiAA), minimum height = 3.4cm] at (3,-1) (xi) {};
\node [slantedbox = $\xi_*$, fit = (phi*QA)(phi*Q*A)(phi*AA), minimum height = 3.4cm] at (4,-5.75) (xi*) {};
\node [slantedbox = $\xi_{**}$, fit = (phi**AA), minimum height = 2.6cm] at (4.5,-10) (xi**) {};
\end{pgfonlayer}

\end{tikzpicture}
\end{center}
\caption{The flow diagram for the random clustered networks.}
\label{fig:phi_xis}
\end{figure}


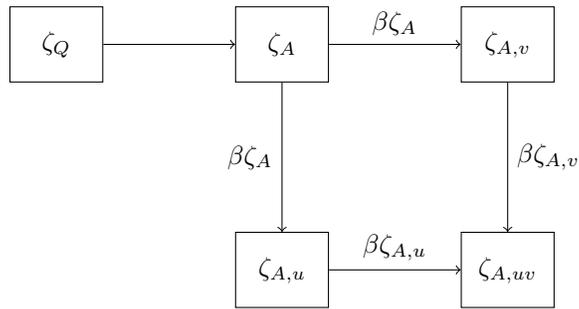
\begin{figure}
\begin{center}
\begin{tikzpicture}
\node [box] (zetaQ) at (0,0) {$\zeta_Q$};
\node [box] (zetaA) at (3,0) {$\zeta_A$};
\node [box] (zetaAv) at (6,0) {$\zeta_{A,v}$};
\node [box] (zetaAu) at (3,-3) {$\zeta_{A,u}$};
\node [box] (zetaAuv) at (6,-3) {$\zeta_{A,uv}$};
\path [->,left] (zetaA) edge node {$\beta\zeta_A$} (zetaAu);
\path  [->, above] (zetaAu) edge node  {$\beta\zeta_{A,u}$} (zetaAuv);
\path [->,above] (zetaA) edge node  {$\beta\zeta_A$} (zetaAv);
\path [->,right] (zetaAv) edge node {$\beta\zeta_{A,v}$} (zetaAuv);
\path [->] (zetaQ) edge node {} (zetaA);
\end{tikzpicture}
\end{center}
\caption{Flow diagram for the $\zeta$ variables.  }
\label{fig:zetas}
\end{figure}

We again take $u$ to be a test node.  We consider a triangle
involving $u$, $v$, and $w$.  Taking $\xi(t)$, $\xi_*(t)$, and
$\xi_{**}(t)$ to be the probabilities that the neighbors in the
triangle have combined to transmit $0$, $1$, or $2$ times to $u$ we
have
\[
Q(t) = \sum_{r>0} \sum_{m=0}^{r-1} \left(\frac{\psi_{r,\triangle}^{(m)}(\xi)}{m!}
\sum_{d=0}^{r-m-1} \binom{m}{d}\xi_*^{m-d} \xi_{**}^d\right) \, .
\]
To interpret this equation, note that the sum over $r>0$ considers all
thresholds for nodes which are not initially active.  The sum
over $m$ represents the number of triangles that have transmitted to
$u$.  We have $\psi_{r,\triangle}^{(m)}(\xi)/m!$ is the probability
that a node has a given $r$ and has had exactly $m$ triangles
transmit.  Of those $m$ triangles we take $d$ to be the number which
have transmitted twice.  So given $m$ and $d$ a node has
received $m+d$ transmissions.  So long as $d$ is at most $r-m-1$,
$m+d$ is at most $r-1$.  The probability of a given $m$ and $d$
occurring is $\binom{m}{d} \xi_*^{m-d} \xi_{**}^d$.

Once we have $Q$ we find
\[
A = 1-Q \, .
\]
That is if an individual is not counted in the $A$ class it is considered in the $Q$ class.

To calculate the $\xi$ variables, we will need to know $\delta_0$, the
probability that $w$ is quiescent given that $u$ and $v$ are prevented
from transmitting to $w$, and $\delta_1$, the probability that $w$
would still be quiescent even if $v$ has transmitted to it (and $u$ has
not).  These are
\begin{align*}
\delta_0
&= \frac{1}{\ave{K_\triangle}}\sum_{r>0} \sum_{m=0}^{r-1}
\left(\frac{\psi_{r,\triangle}^{(m+1)}(\xi)}{m!} \sum_{d=0}^{r-m-1}
\binom{m}{d} \xi_*^{m-d}\xi_{**}^d\right)\\
 \delta_1 
&= \frac{1}{\ave{K_\triangle}}\sum_{r>0} \sum_{m=0}^{r-2}
\left(\frac{\psi_{r,\triangle}^{(m+1)}(\xi)}{m!}
\sum_{d=0}^{r-m-2}\binom{m}{d} \xi_*^{m-d}\xi_{**}^d\right) \, .
\end{align*}
We now introduce some auxiliary variables which help to find the
$\xi$ variables.  We assume $u$ and $v$ are prevented from
transmitting to $w$.  We define $\zeta_Q$ to be the probability
that $w$ is quiescent, $\zeta_A$ the probability it is active but
has not transmitted to $u$ or $v$, $\zeta_{A,u}$ the probability it
has transmitted to $u$ but not $v$, $\zeta_{A,v}$ the probability it
has transmitted to $v$ but not $u$, and $\zeta_{A,uv}$ the probability
it has transmitted to both.  Figure~\ref{fig:zetas}
demonstrates the flow of these variables.  We have
\begin{align*}
  \zeta_Q &= \delta_0 \\
  \dot{\zeta}_{A,u} &= \beta \zeta_A - \beta \zeta_{A,u}\\
  \dot{\zeta}_{A,v} &= \beta \zeta_A - \beta \zeta_{A,v}\\
  \dot{\zeta}_{A,uv} &= \beta (\zeta_{A,u} +\zeta_{A,v})\\
  \zeta_A &= 1- \zeta_Q - \zeta_{A,u}-\zeta_{A,v} - \zeta_{u,v}\, .
\end{align*}
We now introduce our final set of variables.  We define $\phi_{AB}$ to
be the probability of having one neighbor of status $A$ and the other
of status $B$.  We use $Q^*$ in the subscript of $\phi$ to denote a
quiescent neighbor that has received transmission from the other
neighbor.  We use $*$ or $**$ in the superscript of $\phi$ to denote
that $u$ has received $1$ or $2$ transmissions from the neighbors in
the triangle.  We can find most of these variables in terms of the
$\zeta$ variables.  The others ($\phi_{AA}$, $\phi^*_{AA}$, and
$\phi^{**}_{AA}$) can be found in terms of the $\xi$
variables, following figure~\ref{fig:phi_xis}:
\begin{align*}
\phi_{QQ}(t) &= \delta_0^2 = \zeta_Q^2\\
\phi_{QA}(t) &= 2\delta_0 \zeta_A\\
\phi_{Q^*I}(t) &= 2\delta_1\zeta_{A,v}\\
\phi^*_{QA}(t) &= 2\delta_0 \zeta_{A,u}\\
\phi^*_{Q^*I}(t) &= 2\delta_1 \zeta_{A,uv}\\
\phi_{AA}(t) &= \xi- \phi_{QQ} - \phi_{QA} - \phi_{Q^*I}\\
\phi^*_{AA}(t) &= \xi_* - \phi^*_{QA} - \phi^*_{Q^*I}\\
\phi^{**}_{AA}(t) &= \xi_{**}\, .
\end{align*}

\begin{figure}
\begin{center}
\includegraphics[width=0.45\columnwidth]{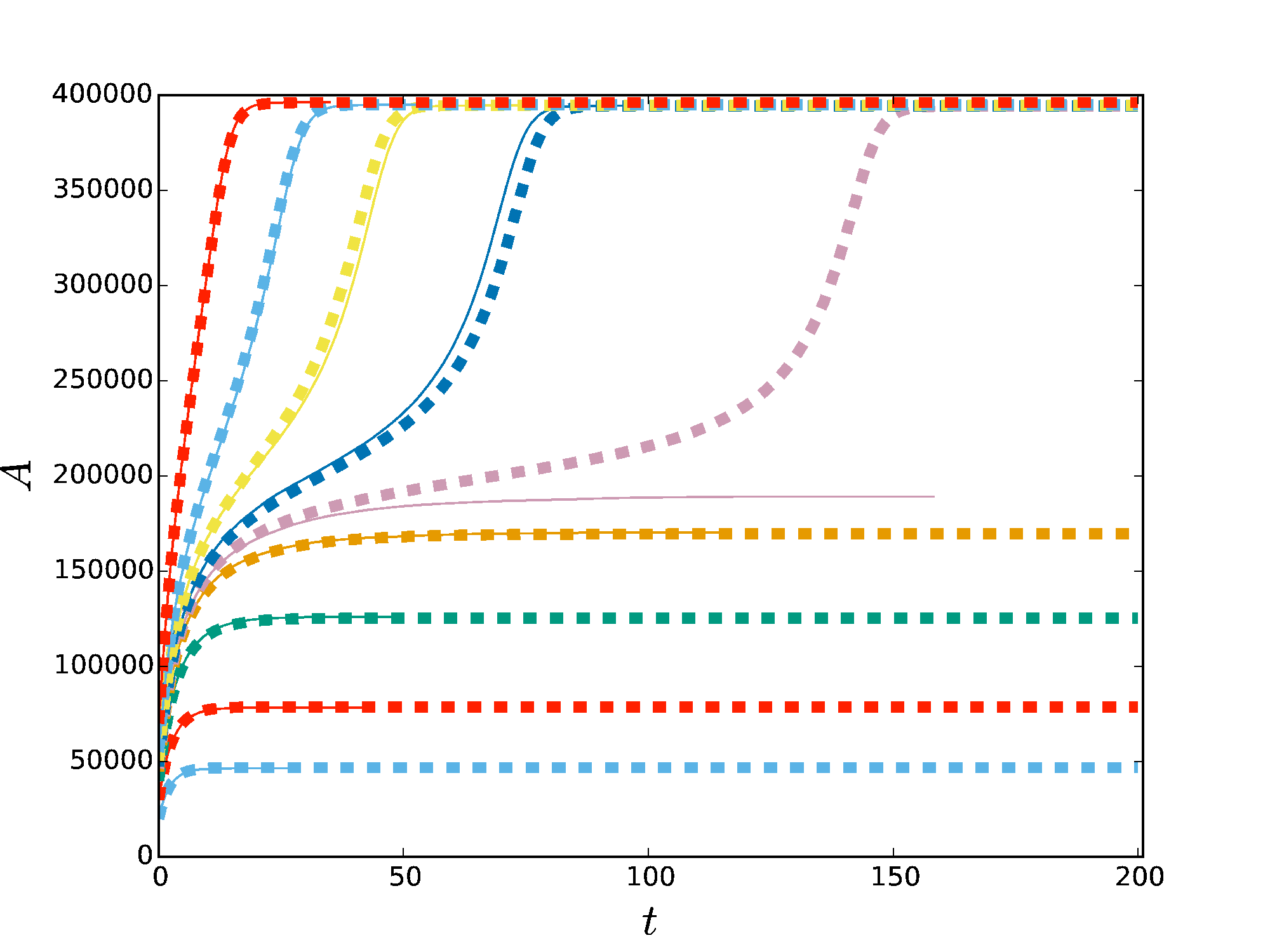}
\hfill\includegraphics[width=0.45\columnwidth]{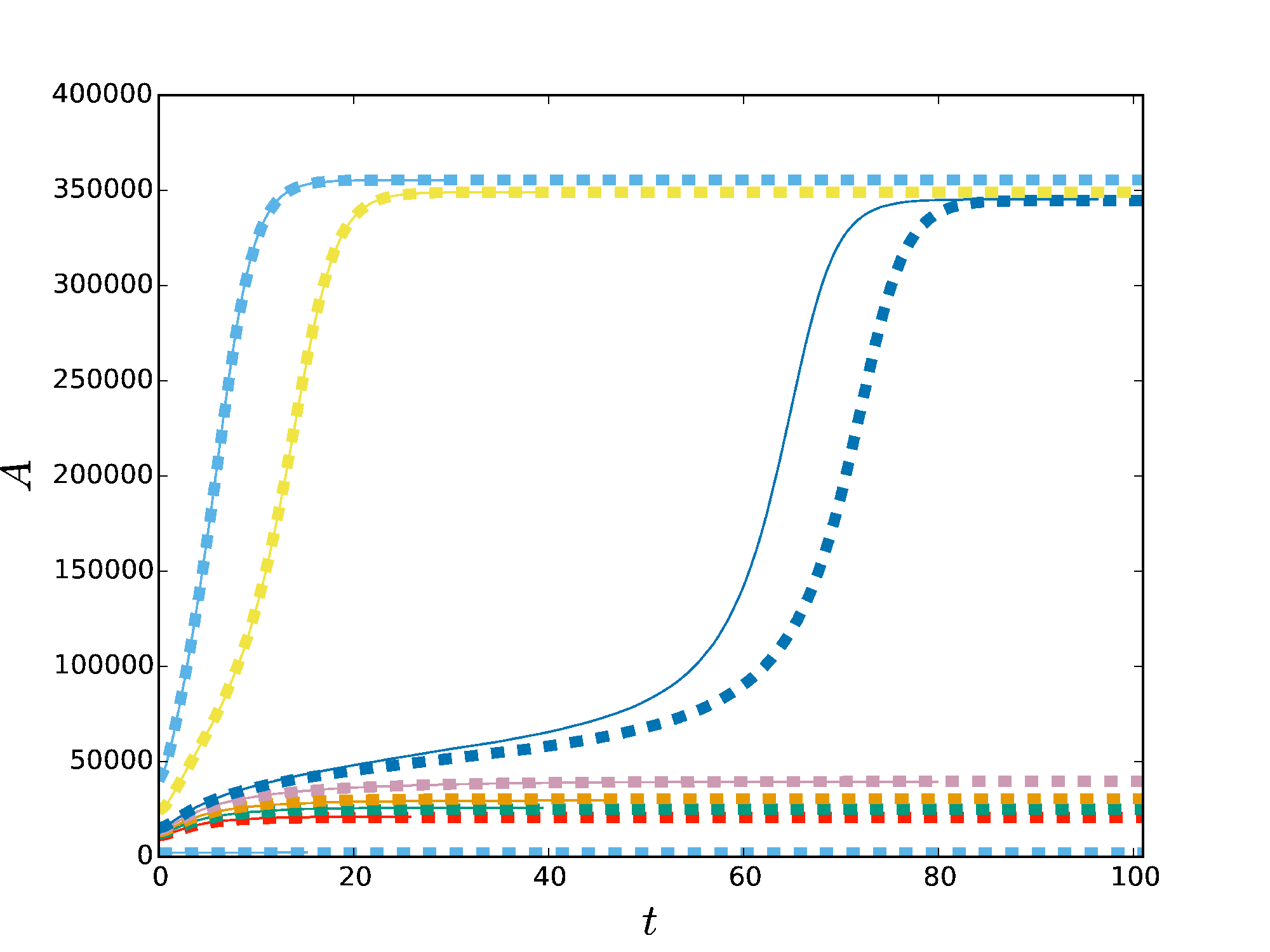}
\end{center}
\caption{  Comparison of simulations (solid) and predictions (dashed) for
  different sized initial conditions in a network with
  $P_\triangle(1,2) = P_\triangle(2,2) = P_\triangle(3,2) = 1/3$
  (left) and $P_\triangle(1,1)=P_\triangle(2,2)=P_\triangle(3,3)= 1/3$
  (right) in populations of $400000$ nodes.  Note that the
  degree of a node is twice its number of triangles $k_\triangle$.  In
  both cases there is a hybrid phase transition occurring for large
  enough initial condition.  In one case, a system predicted to be
  just above the cascade threshold failed to form a cascade due
  to stochastic chance.  Close to this threshold, stochastic effects
  are important and the simulations do not exactly match prediction.
  With larger populations the difference would shrink.}
\label{fig:clusteredsim}
\end{figure}

We can now write down our final differential equations for the
derivatives of the $\xi$ variables
\begin{align*}
\dot{\xi} &= - \beta (\phi_{QA} + \phi_{Q^*I} + 2\phi_{AA})\\
  \dot{\xi}_*&= \beta (\phi_{QA} + \phi_{Q^*I} + 2\phi_{AA}) - \beta
  \phi^*_{AA}\\
  \dot{\xi}_{**} &= \beta \phi^*_{AA}\, .
\end{align*}
We now have a closed system of equations.  We could replace the
$\dot{\xi}_{**}$ equation with $\xi_{**}=1-\xi-\xi_*$.  Solutions to our equations are shown in figure~\ref{fig:clusteredsim}. The generalization of our approach to nonzero $k_I$ would be
straightforward but technical.

\subsection{Discrete time}
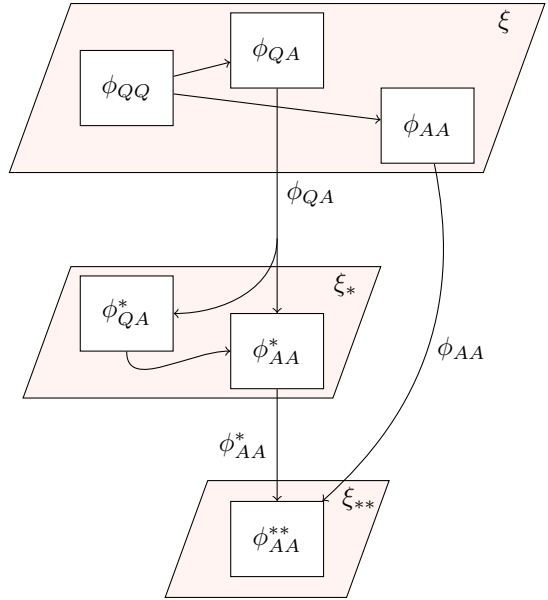
\begin{figure}
\begin{center}
\begin{tikzpicture}
\node [box] (phiQQ) at (0,0) {$\phi_{QQ}$};
\node [box] (phiQA) at (2,0.5) {$\phi_{QA}$};
\node [box] (phiAA) at (4,-0.5) {$\phi_{AA}$};
\node [box] (phi*QA) at (0,-3) {$\phi^*_{QA}$};
\node [box] (phi*AA) at (2,-3.5) {$\phi^*_{AA}$};
\node [box] (phi**AA) at (2,-6) {$\phi^{**}_{AA}$};

\coordinate [] (empty) at (2,-2) {};
\path [->] (phiQQ) edge node {} (phiQA);
\path [->] (phiQQ) edge node {} (phiAA);

\path [-, right] (phiQA) edge node[pos=0.7]  {$\phi_{QA}$} (empty);
\path [->] (empty) edge [out=270, in = 0]  node {} (phi*QA);
\path [->, right] (empty) edge [out=270,in=90]node {} (phi*AA);
\path [->] (phi*QA) edge [out=270,in = 180] node {} (phi*AA);
\path [->, left] (phi*AA)  edge node {$\phi^*_{AA}$} (phi**AA);
\path [->, bend left, right] (phiAA) edge node {$\phi_{AA}$} (phi**AA);
\begin{pgfonlayer}{background}
\node [slantedbox=$\xi$, fit = (phiQQ)(phiQA)(phiAA)] (xi) at (2,0) {};
\node [slantedbox=$\xi_*$, fit = (phi*QA)(phi*AA)] (xi*) at (1,-3.25) {};
\node [slantedbox=$\xi_{**}$, fit = (phi**AA), minimum width = 3cm] (xi**) at (2,-6) {};
\end{pgfonlayer}
\end{tikzpicture}
\end{center}
 \caption{The flow diagram for the discrete time (synchronous update)
   random clustered network model.  At each time step any active node transmits to all of
   its neighbors.  The diagram is somewhat simpler than in the
   continuous time (asynchronous update) case.  Only those edges which transit between $\xi$, $\xi_*$ and $\xi_{**}$ are labeled as they are needed to determine the values of the $\xi$ variables.  The other variables can be determined in terms of the $\zeta$, $\delta$, and $\xi$ variables.}
\label{fig:discrete_clustered}
 \end{figure}

We now consider the discrete time version of the spread in these
triangle-based networks.  The dynamics are simplified because we do not need
to consider cases where a node has transmitted to some, but not all, of
its neighbors.  Figure~\ref{fig:discrete_clustered} shows how the model changes for the discrete case.

The non-ODE equations remain the same except that
$\zeta_{A,u}=\zeta_{A,v}=0$, which implies
$\phi_{Q^*I}=\phi^*_{QA}=0$.  The ODEs are replaced.  We have $\zeta_A(t) =
1-\delta_0(t) - \zeta_{A,uv}(t)$ 
and
\begin{align*}
\xi(t+1) &= \xi(t) - \phi_{QA}(t) - \phi_{AA}(t)\\
\xi_*(t+1) &= \xi_*(t) + \phi_{QA}(t) - \phi^*_{AA}\\
\zeta_{A,uv}(t+1) &= 1-
\delta_0(t) \, .
\end{align*}
Once we have $\zeta_{A,uv}(t+1)$, $\xi(t+1)$, and $\xi_*(t+1)$ we can
build up all of the other $\phi$ variables at time $t+1$.  Then we can
update for the next time.

\subsection{Threshold}
To understand the threshold, the initial analytic method presented for the configuration model case becomes more complicated, so we consider just the second method, calculating the probability that the initial node manages to activate others arbitrarily far from it.  Our result shows that global cascades are possible from arbitrarily small initial conditions ($\rho \to 0$).  This shows that the conjecture of~\cite{watts:WTM} that global cascades only occur if $r=1$ nodes percolate does not generalize to networks with clustering.  This is not a particular surprise: previous researchers have seen this sort of behavior : by activating a single node and all of its neighbors (using ``cluster seeding'') they can initiate a global cascade~\cite{centola:cascade}.  An analytic result of~\cite{hackett:cascades_random_clustered} used a different approach to derive an equivalent threshold to ours for these networks from which the contradiction can also be deduced.

Consider a triangle containing a single active node $u$ early in the spread and determine the probability that the activation results in $1$ or $2$ further transmissions to others $v$ and $w$ in the triangle.  It will transmit to both of them directly, but it will only directly cause activation if $r=1$.  A neighbor has $r=1$ with probability $q_1=\sum_{k_\triangle} k_\triangle P(k_\triangle,1)/\ave{K_\triangle} = \psi_{1,\triangle}'(1)/\ave{K_\triangle}$.  This results in an additional $k (k-1) P(k,1)/\sum kP(k,1) = \psi_{1,\triangle}''(1)/\psi_{1,\triangle}'(1)$ triangles with disease introduced for each $r=1$ neighbor.  So from the two other nodes in the triangle, there are an expected $2q_1 \psi_{1,\triangle}''(1)/\psi_{1,\triangle}'(1) = 2\psi_{1,\triangle}''(1)/\ave{K}$ new triangles with a single introduced activation because of the nodes having $r=1$.

However, there is an additional case to consider.  If one neighbor has $r=1$ and the other has $r=2$, then the $r=2$ node will become active.  Early in the spread, an $r=2$ neighbor will eventually become active from transmissions within the triangle iff the other neighbor has $r=1$.  Thus the probability that one node has $r=2$ and becomes active is the probability that one has $r=2$ and the other has $r=1$.  That is, $2q_1 q_2$ where $q_2 = \sum_{k_\triangle}k P(k_\triangle,2)/\ave{K_\triangle}$.  The resulting number of triangles from the $r=2$ node is $\sum_{k_\triangle} k_\triangle (k_\triangle-1) P(k_\triangle,2)/\ave{K_\triangle}$.  So the expected number of new triangles coming through $r=2$ nodes is $2q_1\psi_{2,\triangle}''(1)/\ave{K_\triangle} = 2 \psi_{1,\triangle}'(1)\psi_{2,\triangle}''(1)/(\ave{K_\triangle}^2$.

So the epidemic threshold occurs when
\[
2 \frac{\psi_{1,\triangle}''(1)}{\ave{K_\triangle}}  + 2
\frac{\psi_{1,\triangle}'(1)\psi_{2,\triangle}''(1)}{\ave{K_\triangle}^2}
=1 \, ,
\]
with a population-scale cascade possible if the left hand side is
larger.  If $r$ values are assigned independently of $k$, this reduces
to 
\[
2
P(r=1)[1+P(r=2)]\frac{\ave{K_\triangle^2-K_\triangle}}{\ave{K_\triangle}}
= 1 \, .
\]
This condition is consistent with that found
by~\cite{hackett:cascades_random_clustered}.

\subsubsection{Condition for hybrid phase transition}
Because of the added variables, it is much more difficult to prove the
existence of an ``interior'' fixed point and the saddle-node
bifurcation.  We will not attempt to prove as strong of a result here.
We anticipate that the conditions for the hybrid phase transition will
be similar.

We assume that there is no cascade in the $\rho \to 0$ limit.  We
consider a configuration model network for which $k$ has the same
distribution as $k_\triangle$.  The active proportion in the
configuration model is less than that of the corresponding triangle
network.  This is because being joined to two nodes in a triangle
increases the probability that at least one transmission has come, and
it also opens the possibility of two transmissions.  If this
configuration model network has a hybrid phase transition, then the
triangle-based network must as well.  This is a fairly crude bound.

\subsection{Impact of clustering}

To investigate the role of clustering, we want to compare the spread
in triangle-based networks and configuration model networks with the same
degree distribution.  Note that in a triangle-based network, a node with a
given $k_\triangle$ has degree $2k_\triangle$, so in the configuration
model networks, $k$ has the same distribution as $2k_\triangle$ in the
triangle-based networks.  We find $\psi_{r,\text{CM}}(x) =
\psi_{r,\triangle}(x^2)$.  

The condition for a cascade from arbitrarily small $\rho$ in a
configuration model network is that
$\psi_{1,\text{CM}}''(1)/\ave{K}>1$.  In this case, we have a cascade
in the Configuration Model network
if 
\begin{align*}
1&< \psi_{1,\text{CM}}''(1)/\ave{K} \\
&= \frac{2\psi_{1,\triangle}'(1) +
4\psi_{1,\triangle}''(1)}{2\ave{K_\triangle}}\\
 &= \frac{\psi_{1,\triangle}'(1)
+ 2\psi_{1,\triangle}''(1)}{\ave{K_\triangle}}\, .
\end{align*}
If the initially active nodes are randomly chosen,
$\psi_{1,\triangle}'(1) = P(r=1)\ave{K_\triangle}$ and
$\psi_{2,\triangle}''(1) = P(r=1) \ave{K_\triangle^2-K}$.  So this becomes
\[
1 < 2P(r=1)\frac{\ave{K_\triangle^2-K}}{\ave{K}}  + P(r=1) \, .
\]
If we instead consider the threshold for a corresponding triangle-based
network, it is
\[
1< 2 P(r=1)\frac{\ave{K_\triangle^2-K_\triangle}}{\ave{K_\triangle}} +
2 P(r=1)P(r=2) \frac{\ave{K_\triangle^2-K_\triangle}}{\ave{K_\triangle}}\, .
\]
If $P(r=2)=0$, then cascades from arbitrarily small initial conditions
are inhibited in the triangle-based network compared to the configuration
model.  However, if $2P(r=2)
\ave{K_\triangle^2-K_\triangle}/{\ave{K_\triangle}}> 1$, cascades from
arbitrarily small initial conditions are enhanced in the triangle-based
network in the sense that the threshold is reduced.

We can interpret this result physically by considering an $r=1$ node $u$ which receives transmission from a neighbor $v$ early in the spread.  We assume $u$ and $v$ share a common neighbor $w$, and we are guaranteed that $v$ will transmit to $w$.  In the absence of $r=2$ nodes, either the $v$ transmission will infect $w$ or the combined $u$ and $v$ transmissions are not enough to infect $w$.  At early time, the $u$-$w$ edge thus has no impact on the spread regardless of $w$'s threshold.  So for initiating a cascade from a small $\rho$ it is irrelevant.  In contrast in an unclustered network, if $u$ has the same degree, then it will have one more neighbor it can transmit to, which may have $r=1$.

\subsubsection{A simple example where clustering inhibits spread}
We can find a simple case where clustering inhibits cascade spread.
If we set $r_u=k_u-1$ for every node, then no triangle can be invaded.
More generally if a network has a $k_0$--clique and every node in this
clique has $r_u > k_u - k_0+1$, then the clique cannot be invaded.  To see this, assume $u$ is the first node in the clique to activate. It has $k_0-1$ neighbors in the clique that we are sure are not yet active, so at most $k_u-k_0+1$ of its neighbors are active.  This contradicts the assumption that it activates.

\section{Discussion}

We have investigated the spread of complex contagions through static networks, focusing on the Watts Threshold Model (WTM).  We have adapted the Edge-Based Compartmental Modeling approach from SIR disease modeling to study the dynamic spread of these infectious processes.  This leads to a compact system of equations for the spread through Configuration Model and triangle-based networks (as introduced by Newman~\cite{newman:cluster_alg} and Miller~\cite{miller:random_clustered}). This framework helps us to unify many disparate results about this and related models under a common framework, and derive some new results about threshold conditions for cascades.  

In Configuration Model networks we find that activation of an arbitrarily small initial proportion can lead to activation of a large fraction of the population iff those nodes who require only a single transmission to become active (\emph{i.e.}, $r=1$) form a giant component (confirming a hypothesis of~\cite{watts:WTM} for this case).  In triangle-based networks $r=2$ nodes can contribute to the initiation of cascades from a small initial active proportion (showing that the hypothesis is not true in the presence of short cycles).

For configuration model networks, we have found a sufficient condition
for a hybrid phase transition to occur when there is a sufficiently
large initially active population.  If
\begin{enumerate}
\item No node has an activation threshold $r_u \geq k_u$
\item $r=1$ nodes do not form a giant component, and
\item $r=k-1$ nodes do not form a giant component,
\end{enumerate}
then there is a threshold concentration of initially active
nodes which will lead to a cascade.

\subsection{Limitations}
The triangle-based networks we study allow for analytic methods.  However,
it comes at a price that the structure of the networks is restricted.
Triangles that share edges are very rare.  If we consider $k$-clique
percolation~\cite{palla:kclique} in other clustered networks and find a
large component and $r\leq k-1$ for all nodes in this component
then if one $k$-clique in the component becomes fully active, the
process will spread throughout the component.  The ``cluster-seeding''
which is often used in simulations where a single node and all of its
neighbors are activated for the initial condition will lead to this
sort of behavior.  This process will not be captured by our random
clustered network model.  In essence, some of the behavior of complex
contagions in more general clustered networks is believed to be a
consequence of ``wide'' bridges.  This is missing in our clustered
networks.

\subsection{Possible extensions}
It is easy to adapt this model to the case where initial active nodes are
targeted by degree.  This simply involves modifying the choice of
$P(k,r)$, noting that $r=0$ corresponds to the initially active nodes.
This appears as a change in $\psi_{r,\text{CM}}(x)$, but we then
cannot assume that $\psi$ is proportional to $1-\rho$.  

We can also adapt this to account for biased mixing, where the degree
of a node provides information about the degree of its neighbors.   We
assume we know $P_n(k'|k)$, that is the probability a neighbor has
degree $k'$ given that a node has degree $k$.  We define $\theta_k$ to
be the probability that a degree $k$ node's neighbor has not yet
transmitted to it.  Our function $\psi$ would then become a summation
of $P(k) \theta_k^k$ than a function of a single variable $\theta$.
Closely related models have been studied by~\cite{melnik:modularnetworks}

We could also adapt this to account for a network which changes in
time.  Assuming for example that existing edges end at some
rate $\eta_1$, and then the freed stubs seek out new neighbors at
another rate $\eta_2$, the methods of~\cite{miller:ebcm_overview} lead
us to the new model equations.  A particularly interesting limit of
this would have $\eta_2 \to \infty$ such that nodes immediately
find new neighbors when their old edges end.

It would be straightforward to add a recovered class to this model,
much as has been done previously in SIR disease models.  Some of the
relevant calculations have already been performed
by~\cite{shrestha:message}.  We could also consider the possibility of
nodes transmitting multiple times.  This would involve
subdividing $1-\theta$ into more components based on the number of
transmissions that have occurred, much like the $\xi_*$ and $\xi_{**}$
components that occurred in the triangle-based network model.

We can adapt this approach to consider multiple competing processes spreading as
done by~\cite{miller:compete, karrer:competing,newman:threshold}.
May be interesting for political opinion.  For understanding the
formation of political groups, it would be interesting to
consider 4 beliefs spreading: $A_1$, $A_2$, $B_1$, and $B_2$ where the
two $A_i$ are incompatible and the two $B_i$ are incompatible, while
the $A$ and $B$ processes are independent, with the population
rewiring and preferentially selecting neighbors with at least one
common belief.

\section*{Acknowledgments}
Mason Porter and James Gleeson provided comments on a primitive version and two anonymous reviewers gave feedback.  This significantly improved the quality of this work.
\providecommand{\noopsort}[1]{}

\end{document}